\documentclass[12pt,a4paper]{article}
\usepackage{amsfonts}
\usepackage{smart1}
\usepackage{latexsym}
\usepackage{Bourbaki}

\newcommand{\sectionnew}[1]{\section{#1}}
\link{equation}{section}\toheight1  

\newcommand{\beq}{\begin{equation}}
\newcommand{\eeq}{\end{equation}}
\newcommand{\beqa}{\begin{eqnarray}}
\newcommand{\eeqa}{\end{eqnarray}}
\newcommand{\nn}{\nonumber \\}

\def \R {{\mathbb R}}
\def \C {{\mathbb C}}
\def \Z {{\mathbb Z}}
\def \N {{\mathbb N}}
\def \S {\mathcal{S}}
\def \Sr {{\mathbb S}}
\def \M {\overline{M}}
\def \SMB {\mathfrak{T}}
\def \Bb {{\mathcal B}}
\def \BB {{\mathbb B}}
\def \Ii {{\mathcal I}}

\def \lvac {\left\langle 0 \! \left| \right. \right.  \!}
\def \rvac {\!\! \left. \left. \right| \! 0 \right\rangle}

\def \di {\partial}

\def \su {{{{{{{{{\sum^{
\mathop{}\limits^{\mathop{}\limits^{}}}
}_{}}_{}}_{}}_{}}_{}}_{}}_{\!\!\!\!} \,}}

\def \Su {\mathop{\sum}\limits}
\def \vspe {\mathop{}\limits_{}^{}}
\def \dvspe {\mathop{}\limits_{}}
\def \gvspe {\mathop{}\limits^{}}
\def \mop {\mathop{}\limits}
\def \dti {{\!\!\!\!\!{}_{{}_{{}_{{}_{}}}}}_{\sim}
{\!{}_{{}_{}}}}
\def \wti {\widetilde}

\def \lb {\left(}
\def \rb {\right)}
\def \Lb {\left[}
\def \Rb {\right]}
\def \LB {\left\{}
\def \RB {\right\}}
\def \l. {\left.}
\def \r. {\right.}
\def \la {\left\langle}
\def \ra {\right\rangle}
\def \l| {\! \left| \,}
\def \r| {\right|}

\def \W {{\cal W}}
\def \WW {{\cal W}}
\def \DD {{\cal D}}
\def \LL {{\cal L}}

\advance\topmargin-2cm 

\begin{document}

\title{
Four Dimensional CFT Models with
Rational Correlation Functions\footnote{Vienna preprint, ESI 1094 (2001)}
}

\author{
N.M. Nikolov{\normalsize ${\!}{\,}^{{\,}^{a)}}$}\footnote{
mitov@inrne.bas.bg}
\ , \quad
Ya.S. Stanev{\normalsize ${\!}{\,}^{{\,}^{a)\, b)}}$}\footnote{
stanev@roma2.infn.it}
\ , \quad
I.T. Todorov{\normalsize ${\!}{\,}^{{\,}^{a)\, c)}}$}\footnote{
todorov@inrne.bas.bg{\,}, \ \, itodorov@esi.ac.at}
\\
\\
{\normalsize
\(
\begin{array}{l}
{\!\!\!\!\!}{\,}^{a)} \,
\mathrm{
Institute \ for \ Nuclear \ Research \ and \
Nuclear \ Energy,
}
\\ \,
\mathrm{
72 \ Tsarigradsko \ Chaussee, \ BG \!\! - \!\! 1784 \
Sofia, \ Bulgaria
}
\\
\\
{\!\!\!\!\!}{\,}^{b)} \,
\mathrm{
Dipartimento \ di \ Fisica, \ Universita \ di \ Roma \
'Tor \ Vergata', \
}
\\ \,
\mathrm{
I. \, N. \, F. \, N. \, - \ Sezione \ di \ Roma \
'Tor \ Vergata', \
}
\\ \,
\mathrm{
Via \ della \ Ricerca \ Scientifica \ 1, \ I - 00133 \
Roma, \ Italy
}
\\
\\
{\!\!\!\!\!}{\,}^{c)} \,
\mathrm{
Erwin \ Schr\ddot{o}dinger \ International \ Institute \ for \
Mathematical \ Physics,
}
\\ \,
\mathrm{
Boltzmanngasse \ 9, \ A \!\! - \!\! 1090 \ Wien, \
Austria
}
\\
\\
\end{array}
\)
}}

\maketitle


\begin{abstract}

Recently established rationality of correlation functions
in a globally conformal invariant quantum field theory
satisfying Wightman axioms is used to construct a family
of soluble models in $4$-dimensional Minkowski space-time.
We consider in detail a
model of a neutral scalar field $\phi$ of
dimension $2\,$. It depends on a positive real parameter $c$,
an analogue of the Virasoro central charge, and admits for all
(finite) $c$ an infinite number of conserved symmetric tensor
currents.
The operator product algebra of $\phi$ is shown to
coincide with a simpler one, generated by a bilocal scalar
field $V \left( x_1,\, x_2 \right)$ of dimension
$\left( 1,\, 1 \right)\,$.
The modes of $V$ together with the unit operator span an
infinite dimensional Lie algebra $\mathfrak{L}_V$ whose
\textit{vacuum} (i.e. zero energy lowest weight)
representations only depend on the central charge $c\,$.
Wightman positivity (i.e. unitarity of the representations
of $\mathfrak{L}_V\,$) is proven to be equivalent to
$c \in \N\,$.

\end{abstract}



\vspace{0.1in}


\vspace{0.15in}

\pagebreak 



\section{Introduction}

The task of constructing a conformally invariant quantum
field theory model - using dressed vertices and (global)
operator product expansions (OPE) - has been set forth over
30 years ago
(\cite{Pol70} \cite{Mig71} \cite{PP71}
\cite{ESch71} \cite{BSch71-73} \cite{MS72}
\cite{Sym72} \cite{FGGP72-73} \cite{Tod72}
\cite{MT73} \cite{Mack73-74} \cite{Pol74}
\cite{SSV75} \cite{DPPT76} \cite{Mack77a};
for a review of this early work and further references - see
\cite{TMP78}).
After a relative quiet
(during which only some sporadic applications of the formalism
appeared - see e.g.
\cite{CDT85})
the subject has been gradually revived
(see \cite{Tod86} \cite{Car87} \cite{Sta88} \cite{LR93}
\cite{OP94} \cite{FP98} \cite{DO 01}
among others) in the wake of the $2$--dimensional (2D)
conformal field theory (CFT) revolution (now the subject of
textbooks-- see e.g.
\cite{DMS96}
where a bibliography of original work can be found).
It gathered new momentum with the discovery of the AdS--CFT
correspondence and the associated intensified study of the
$N=4$ supersymmetric Yang-Mills theory (for a sample of recent
papers and further references - see
\cite{AEPS 01} \cite{BKRS 01}).

The present work is chiefly motivated by the concept of a
rational conformal field theory (RCFT). Albeit this notion
arose in the framework of 2D CFT, recent work
\cite{NT 01}
suggests that it may be relevant to any number of space--time
dimensions.
We consider in detail
the simplest example beyond free
fields, given in \cite{NT 01},
the case of a
model of a
neutral scalar field of dimension $2\,$.
More complicated (and potentially more interesting)
cases involving fields of dimension $3$ and $4$ are
only briefly discussed.

We start by recalling the relevant results of
\cite{NT 01}
which allow us to derive the general expressions for the
$4$--point Wightman functions.

Adding to the Wightman axioms a condition of global conformal
invariance
(GCI)
of local observables (i.e. invariance of
correlation functions under a single--valued action of the
$4$--fold cover $G=\mathrm{SU}\lb  2,2\rb$ of the conformal
group whenever $x$ and $gx$ ($g \in G$) both belong to
Minkowski space) we deduce
the Huygens' principle:
local fields $\phi \left( x \right)\,$,
$\psi \left( y \right)$ commute whenever the difference
$x-y$ is non--isotropic;
moreover,
\beq\label{new1}
\left[ \left( x-y \right)^2 \right]^N \,
\left[ \, \phi \left( x \right) \, , \,
\psi \left( y \right) \, \right]
\, = \, 0
\quad \mathrm{for} \quad N \, > \!\! > \, 1
\qquad
\eeq
(see \cite{NT 01} Theorem 4.1 and Proposition 4.3,
where the precise bound for $N$ is given).
This result is based on the fact that a space--like separated pair of points in
Minkowski space can be mapped by a proper conformal transformation into a
time-like one.
(Thus, GCI is a stronger requirement than invariance of
Schwinger functions under the euclidean conformal group.)
The Huygens' principle implies (together with energy positivity)
that the Wightman distributions
are rational functions of the form
\beq\label{1.1}
\W \lb x_1,...,\, x_n \rb
\ \lb \ \equiv \ \la 1, \, ... \ n \ra \rb
\ = \
P \lb x_1,...,\, x_n \rb \
\mathop{\prod}\limits_{1 \, \leq \, j \, < \, k \, \leq \, n}
\ \lb \rho_{jk} \rb^{-\mu_{jk}} \quad , \qquad
\eeq
where $P$ is a polynomial
(in general, tensor valued),
\beq\label{1.2}
x_{jk} \ \equiv \ x_j \, - \, x_k
\quad , \quad
\rho_{jk} \ = \ x_{jk}^{\ 2} \, + \, i0\, x_{jk}^0
\quad \lb x^2 \, = \, \mathbf{x}^2 \, - \, x_0^2 \rb
\ , \quad \mu_{jk} \, \in \, \Z_+
\qquad
\eeq
(\cite{NT 01}
Theorem 3.1;
the $i0x^0_{jk}$
is only essential when $\rho_{jk}$ occurs in denominators and
prescribes the contour integration for Wightman distributions that
reflects energy positivity - see \cite{SW64-00}).
Hilbert space positivity is taken
into account using OPE and the classification of positive
energy unitary irreducible representations of $G$
(\cite{Mack77b}).

Expanding the discussion of Sec. 5 of \cite{NT 01} we
shall derive the general form of the truncated $4$--point
function ${\cal W}_{\, 4}^{\, t} \lb d \rb$
of a neutral scalar field $\phi$ of integer
dimension $d$ satisfying GCI
(see Eq. (\ref{r2.1}) below).

Combining Proposition 5.3 and Corollary 4.4 of \cite{NT 01}
we can write
\beqa\label{1.3}
&&
{\cal W}_{\, 4}^{\, t} \lb d \rb \, \equiv \,
{\cal W}^{\, t} \lb x_1,\, ...,\, x_4;\, d \rb \ = \
{\cal D}_d \lb \rho_{ij} \rb \,
{\cal P}_d \lb \eta_1,\, \eta_2 \rb
\ , \quad
\nn &&
{\cal D}_d \lb \rho_{ij} \rb \ = \
\frac{\lb \rho_{13} \, \rho_{24} \rb^{d-2}}{
\lb \rho_{12} \, \rho_{23} \, \rho_{34} \,
\rho_{14} \rb^{d-1}}
\ , \quad
{\cal P}_d \lb \eta_1,\, \eta_2 \rb \ = \
\Su_{\mop^{i,\, j \, \geq \, 0}_{
i+j \, \leq \, 2d - 3}}
\!
c_{ij} \, \eta_1^i \, \eta_2^j
\ , \qquad
\eeqa
where $\eta_i$ are the conformally invariant cross
ratios
\beq\label{1.4}
\eta_1 \ = \ \frac{\rho_{12} \, \rho_{34}}{
\rho_{13} \, \rho_{24}} \ , \quad
\eta_2 \ = \ \frac{\rho_{14} \, \rho_{23}}{
\rho_{13} \, \rho_{24}} \ . \qquad
\eeq
For $x_{jk}^{\ 2} \neq 0$ we can ignore the
$i0x_{jk}^0$ term in the definition of $\rho_{jk}$
(\ref{1.2}).
The Huygens' principle (strong locality) then implies
symmetry under the permutation group $\S_4\,$.
Its normal subgroup $\Z_2 \times \Z_2$
(with non-trivial elements $s_{12}\,s_{34}\,$,
$s_{14}\,s_{23}\,$, $s_{13}\,s_{24}\,$, where
$s_{ij}$ is a substitution exchanging $i$ and $j$)
acts trivially on $\eta_1$ and $\eta_2\,$.
Hence, it suffices to impose invariance under the
$6$--element factor group
$\S_4 \left/ \, \Z_2 \! \times \! \Z_2 \right.
\, \cong \, \S_3$
generated by
\beqa\label{1.5}
&&
s_{12} \, : \ {\cal P}_d \lb \eta_1,\, \eta_2 \rb
\ \mapsto \ \eta_2^{2d-3} \ {\cal P}_d
\lb \frac{\eta_1}{\eta_2},\, \frac{1}{\eta_2} \rb
\ = \ {\cal P}_d \lb \eta_1,\, \eta_2 \rb \ , \quad
\nn &&
s_{23} \, : \ {\cal P}_d \lb \eta_1,\, \eta_2 \rb
\ \mapsto \ \eta_1^{2d-3} \ {\cal P}_d
\lb \frac{1}{\eta_1},\, \frac{\eta_2}{\eta_1} \rb
\ = \ {\cal P}_d \lb \eta_1,\, \eta_2 \rb \qquad
\eeqa
(which also involves
$s_{13} = s_{12} \, s_{23} \, s_{12}$
$= s_{23} \, s_{12} \, s_{23}$
implying
$\, {\cal P}_d \lb \eta_2,\, \eta_1 \rb$
$= {\cal P}_d \lb \eta_1,\, \eta_2 \rb\,$).
This leaves us with the following\footnote{
$\Lb \! \Lb a \Rb \! \Rb$ stands for the integer part of
$a\,$
($\,\Lb \!\! \Lb \frac{d^2}{3} \Rb \!\! \Rb$
$= \, 1$, $3\,$, $5\,$, $8\,$, for $d$ $= \, 2$,
$3\,$, $4\,$, $5\,$;
$\Lb \!\! \Lb \frac{\lb d + 1 \rb^2}{3} \Rb \!\! \Rb$
$- \, \Lb \!\! \Lb \frac{d^2}{3} \Rb \!\! \Rb$
$= \, \Lb \!\! \Lb \frac{2}{3} \ \lb d + 1 \rb \Rb \!\! \Rb\,$
$=\, 1,\, 2,\, 2,\, 3$ for $d$ $=\, 1,\, 2,\, 3,\, 4\,$).}
$\Lb \!\! \Lb \frac{d^2}{3} \Rb \!\! \Rb$
independent coefficients:
\beqa\label{1.6}
&&
c_{i j} \quad \mathrm{for} \quad
i \, \leq \, j \, \leq \,
\frac{2d - 3 - i}{2}
\nn &&
(\, c_{i j} \, = \, c_{j i} \, = \,
c_{i ,\, 2d - 3 - i - j}
\, = \, c_{2d - 3 - i - j ,\, i}
\, = \, c_{j ,\, 2d - 3 - i - j}
\, = \, c_{2d - 3 - i - j ,\, j}
\, )
\ . \qquad
\eeqa

The
present paper is chiefly devoted to the case $d=2\,$, that is the
minimal $d$ for which a non--zero truncated
$4$--point function ${\cal W}_{4}^{\, t} \lb d \rb$
exists.
We shall set in this case\footnote{
The $4$--point Wightman function obtained from (\ref{1.7})
coincides with the one given by Proposition 5.3 and
Eq. (5.16) of \cite{NT 01} for
$N_2$ $= \, \frac{c_2}{32 \, \pi^4}\,$,
$C_2$ $=\, \frac{c_4}{\lb 2 \, \pi \rb^8}\,$,
$C_{20}$ $= \, C_{21}$ $= \, 0$.}
\beqa\label{1.7}
&&
\la 12 \ra \ = \ \frac{c_2}{2} \lb 1 2 \rb^2
\ , \quad
\la 1 2 3 \ra \ = \ c_3 \, \lb 12 \rb \lb 23 \rb \lb 13 \rb
\ , \quad
\nn &&
{\cal W}_{\, 4}^{\, t} \lb d = 2 \rb \ = \
\ c_4 \, \lb 12 \rb \lb 23 \rb \lb 34 \rb \lb 14 \rb \,
\lb 1 \, + \, \eta_1 \, + \, \eta_2 \rb
\ , \quad
\nn &&
\lb i j \rb \ = \ \lb 4 \, \pi^2 \, \rho_{ij} \rb^{-1}
\ . \qquad
\eeqa
Parameters like
\beq\label{1.8}
c \ := \ \frac{c_2^3}{c_3^2} \ = \ 8 \,
\frac{\la 12 \ra \la 23 \ra \la 13 \ra}{\lb \la 123 \ra \rb^2}
\ , \qquad
c' \, := \, \frac{c_2^2}{c_4}
\eeq
are invariant under rescaling of $\phi\,$.
It will be proven in Sec. 2 that if there is a single field
($\,\phi\,$) of dimension $2$ then these constants are equal.
Moreover, their common value $c$ ($\, = \, c'\,$) also
determines the normalization of the $2$--point function of
the stress--energy tensor and thus appears as a generalization
of the Virasoro central charge.
We will then restrict our attention to the case of a single
field $\phi$ corresponding to $c_2$ $= \, c_3$ $= \, c_4$
$= \, c\,$.

Similarly, the
general truncated $4$--point  function for $d=3$ is
\beqa\label{1.9}
{\cal W}_{\, 4}^{\, t} \left( 3 \right) \, = \,
\frac{\rho_{13} \, \rho_{24}}{
\lb \rho_{12} \, \rho_{23} \, \rho_{34} \,
\rho_{14} \rb^2} \
&& \!\!\!\!\!\!\!\!\!\!\!
\LB \vspe \!
c_0 \lb 1 + \eta_1^3 + \eta_2^3 \rb
+ c_1 \Lb
\lb \eta_1 + \eta_2 \rb
\lb 1 + \eta_1 \, \eta_2 \rb +
\eta_1^2 + \eta_2^2
\Rb
+ \right.
\nn && \!\!\!\!\!\!\!
\left.
+ \ b \, \eta_1 \, \eta_2
\vspe \! \RB
\nn && \!\!\!\!\!\!\!
\!\!\!\!\!\!\!\!\!\!\!\!\!\!\!\!\!\!\!\!\!\!\!\!\!\!\!
\!\!\!\!\!\!\!\!\!\!\!\!\!\!\!\!\!\!\!\!\!\!\!\!\!\!\!
\!\!\!\!\!\!\!\!\!\!\!
(\,
c_i \, \equiv \, c_{0i} \quad \mathrm{for} \quad
i \, = \, 0,\, 1 \, , \
\quad \mathrm{and} \quad
b \, \equiv \, c_{11}
\, )
\ . \qquad
\eeqa
The requirement that no $d=2$ (scalar) field is present
in the OPE of two $\phi$'s in this case gives
$c_1 = - c_0$ ($\neq 0\,$, should one demand the
presence of a stress energy tensor in the OPE).

The case $d=4$ appears to be particularly interesting
and will be briefly discussed in the concluding Sec. 6.

The paper is organized as follows.

In Sec. 2
we write down the OPE of two $\phi$'s in terms of a bilocal
scalar field $V \left( x_1,\, x_2 \right)$
of dimension $\left( 1,\, 1 \right)$ which satisfies{\ }--{\ }in
each argument{\ }--{\ }the (free) d'Alembert equation.
Using this result we sketch a proof of the statement that
$V$ belongs to the OPE algebra generated by $\phi\,$,
a property only valid in four space--time
dimensions.
The free field equations for $V$ then imply that the
truncated $n$--point function of $\phi$ is expressed
as a sum of $1$--loop diagrams with propagators
$\left( i j \right)$ and a common factor $c_n\,$
for all $n\geq 4\,$.
The uniqueness of the field $\phi$ of dimension 2 is
proven to correspond to $c_n$ $= \, c \, \alpha^n\,$.

In Sec. 3 we establish the existence of an infinite set
of conservation laws:
the term with light cone singularity
$\lb 12 \rb \lb 34 \rb$ is reproduced by the contribution
of an infinite number of (even rank) conserved symmetric
traceless tensor currents
\beq\label{1.12}
T_{2l} \lb x,\, \zeta \rb \ = \
T_{\mu_1 \, ... \, \mu_{2l}} \lb x \rb \
\zeta^{\mu_1} \! \dots \, \zeta^{\mu_{2l}}
\ , \quad
\Box_{\zeta} T_{2l} \lb x,\, \zeta \rb \ = \ 0 \ = \
\frac{\partial^2}{\partial x_{\mu} \, \partial \zeta^{\mu}}
\ T_{2l} \lb x,\, \zeta \rb \ , \qquad
\eeq
to the OPE of two $\phi$'s
(including the $l=0$ term
$T_0 \lb x \rb = \phi \lb x \rb\,$).
For $\phi$ expressed as a linear combination of normal
products of free fields
\beq\label{1.13}
\phi \lb x \rb \ = \ \frac{1}{2} \,
\Su_{i \, = \, 1}^{N} \, \alpha_i : \!
\varphi_i^2 \lb x \rb \! :
\ , \quad
\la 0 \l|
\varphi_i \lb x_1 \rb \, \varphi_j \lb x_2 \rb
\r| 0 \ra
\ = \ \delta_{ij} \lb 12 \rb
\ \qquad
\eeq
the stress--energy tensor is also given by the sum of free
field expressions:
\beq\label{n1.14}
T_2 \left( x,\, \zeta \right) \, = \,
\Su_{i \, = \, 1}^N \, : \!
\LB
\left( \zeta \! \cdot \!
\partial \varphi_i \left( x \right) \right)^2
- \frac{1}{2} \ \zeta^2 \,
\partial_{\mu} \varphi_i \ \partial^{\mu} \varphi_i +
\frac{1}{6} \ \Lb\zeta^2 \, \Box -
\lb \zeta \! \cdot \! \partial \rb^2 \Rb
\varphi_i^2 \left( x \right) \RB \! :
\ . \qquad
\eeq
The case of equal $c_n$ $=\, c$
($\, n$ $=\, 2\,$, $3\, ...\,$)--
i.e. of a unique $\phi\,$-- corresponds to
$\alpha_i$ $=\, 1$ (for $i$ $=\, 1\, ...\,$, $N\,$) and
$c$ $=\, N\,$.
The truncated $n$--point functions of
$T_2 \left( x,\, \zeta \right)$ remain proportional to its free
massless scalar field expression for all $c > 0\,$.
Thus, the parameter $c$ indeed plays the role of a
$4$--dimensional extension of the Virasoro central
charge.

In Sec. 4 we study the mode expansion of the bilocal
field $V$ which naturally appears in the so called analytic
compact picture.
We exhibit an infinite dimensional Lie algebra
$\mathfrak{L}_V$ spanned by the modes
$V_{nm} \left( z_1,\, z_2 \right)$ of $V$ and by the unit
operator.

In Sec. 5 we prove  that the unitary positive energy
representations of $\mathfrak{L}_V$ correspond to positive
integer $c$ (Theorem 5.1).
Combining this theorem
with Propositions 2.2 and 2.3 we derive
the same result
for the original field algebra of the
$d = 2$ scalar field $\phi\,$.
This implies that
$\phi$ belongs to the Borchers' class
of a system of free fields
\cite{Bor 60}
(see \cite{SW64-00}
for a text-book introduction to this concept).

Sec. 6 is devoted to a discussion of the results.
We indicate
on the way how the methods of this paper apply to
fields of dimension $3$ and $4\,$,
and end up with the formulation of two open problems.





\sectionnew{
One loop $n$--point functions. OPE in terms of a bilocal
field}

We begin by rewriting the expression for the general
$4$--point function of a neutral scalar field
$\phi \lb x \rb$ of dimension $2$
satisfying GCI
in a form that suggests its generalization to the
$n$--point function.
According to (\ref{1.7}) we have
\beq\label{r2.1}
\la 1234 \ra \, = \, \la 12 \ra \la 34 \ra \, + \,
\la 13 \ra \la 24 \ra \, + \, \la 14 \ra \la 23 \ra
\, + \, {\cal W}_{\, 4}^{\, t} \ , \quad
(\, \la ij \ra \, = \, \frac{c_2}{2} \, \lb ij \rb^2 \, )
\ , \qquad
\eeq
where the truncated $4$--point Wightman function
can be written as a sum of contributions of three box
diagrams:
\beq\label{r2.2}
{\cal W}_{\, 4}^{\, t}
\, = \,
\, c_4 \,
\LB
\lb 12 \rb \lb 34 \rb \lb 23 \rb \lb 14 \rb \, + \,
\lb 12 \rb \lb 34 \rb \lb 13 \rb \lb 24 \rb \, + \,
\lb 13 \rb \lb 24 \rb \lb 14 \rb \lb 23 \rb \,
\RB
\ . \qquad
\eeq

This expression is reproduced by an OPE for the product
of two $\phi$'s that can be written compactly in terms
of bilocal fields:
\beq\label{r2.3}
\la 0 \l| \right. \right. \!\!
\phi \lb x_1 \rb \phi \lb x_2 \rb
\, = \,
\la 0 \l| \right. \right.
\!\!\!
\LB \vspe \!
\la 12 \ra +
\lb 12 \rb V \lb x_1, x_2 \rb +
: \! \phi \lb x_1 \rb \phi \lb x_2 \rb \! :
\vspe \! \RB
\, , \,\,\,\,\,
V \left( x_1, x_2 \right) \, = \,
V \left( x_2, x_1 \right)
, \qquad
\eeq
where the three terms are mutually orthogonal
\beq\label{r2.4}
\la 0 \l| \,
V \lb x_1,\, x_2 \rb \, \r| 0 \ra \, = \, 0
\, = \,
\la 0 \l| \, : \! \phi \lb x_1 \rb \phi \lb x_2 \rb \! :
\ \r| 0 \ra
\, = \,
\la 0 \l| \,
V \lb x_1,\, x_2 \rb
: \! \phi \lb x_3 \rb \phi \lb x_4 \rb \! :
\ \r| 0 \ra \ , \qquad
\eeq
and satisfy
\beqa\label{r2.5}
&& \!\!\!\!\!\!\!\!
\la 0 \l| \,
V \lb x_1,\, x_2 \rb V \lb x_3,\, x_4 \rb \ \r| 0 \ra
\, = \, c_4 \,
\LB
\lb 13 \rb \lb 24 \rb  \, + \, \lb 14 \rb \lb 23 \rb
\RB
\ , \quad
\nn && \!\!\!\!\!\!\!\!
\quad \ \ \,
\la 0 \l| \,
V \lb x_1,\, x_2 \rb \phi \lb x_3 \rb \ \r| 0 \ra
\, = \, c_3 \, \lb 13 \rb \lb 23 \rb
\ , \qquad
\\ \label{r2.6} && \!\!\!\!\!\!\!\!
\la 0 \l| \, : \! \phi \lb x_1 \rb \phi \lb x_2 \rb \! :
\phi \lb x_3 \rb \phi \lb x_4 \rb \ \r| 0 \ra \, = \,
\la 13 \ra \la 24 \ra \, + \,
\la 14 \ra \la 23 \ra \, + \,
c_4 \, \lb 13 \rb \lb 23 \rb \lb 14 \rb \lb 24 \rb
\ . \qquad \quad
\eeqa
(In general, a field $V \left( x_1,\, x_2 \right)$
is said to be bilocal if
\(\left[ V \left( x_1,\, x_2 \right),\,
V \left( x_3,\, x_4 \right) \right] = 0\)
for $x_i$ space--like to $x_j\,$,
$i = 1,\, 2\,$, $j = 3,\, 4$ and
if it commutes with all local fields
$\phi \left( x_3 \right)$ of the theory
for space--like \(x_{i3},\, i=1,\, 2\,\).)

A priori, the algebra of $V$ and
$: \! \phi \left( x_1 \right) \phi \left( x_2 \right) \! :$
may be larger than  the OPE algebra of $\phi\,$.
It is a non--trivial result, valid
only in
$4$--dimensions, that the (symmetric) bilocal
fields $V \left( x_1,\, x_2 \right)$
and
$:\! \phi \left( x_1 \right) \phi\left( x_2 \right) \! :$
can actually be determined separately from the
expansion (\ref{r2.3}).
\vspace{0.2in}

\textbf{Proposition 2.1}
\textit{If} $V \left( x_1,\, x_2 \right)$
\textit{is a bilocal field obeying}
(\ref{r2.5}) \textit{then it satisfies
in each argument the d'Alembert equation:}
\beq\label{r2.7}
\Box_1 V \left( x_1,\, x_2 \right)
\, = \, 0 \, = \, \Box_2
V \left( x_1,\, x_2 \right)
\ , \quad
\Box_i \, = \,
\frac{\partial^2}{\partial x_i^{\mu}\,\partial x_{i\,\mu}}
\, , \quad
i \, = \,
1,\, 2
\, , \qquad
\eeq
\textit{provided the metric in the state space is
positive definite.}
\vspace{0.2in}

\textit{Proof.}
The vector valued distribution
$\Box_i V \left( x_1,\, x_2 \right) \rvac\,$,
$i = 1,\, 2\,$, vanishes, due to Wightman positivity
since the norm squares of the corresponding
smeared vectors are expressed in terms of the
$4$--point function in the first equation (\ref{r2.5}).
The vanishing of $\Box_i V$ then follows from
local commutativity by virtue of the
Reeh--Schlieder theorem.
(The argument is essentially the same as the proof of
the statement that the vacuum is a separating vector
for local fields-- see \cite{SW64-00} Sec. 4.)$\quad\Box$
\vspace{0.2in}

\textbf{Proposition 2.2}
\textit{The bilocal field}
\beq\label{r2.8}
W \left( x_1,\, x_2 \right) \, := \,
4 \, \pi^2 \, x_{12}^{\ 2} \,
\left\{
\phi \left( x_1 \right) \phi \left( x_2 \right)
\, - \, \la 12 \ra
\right\}
\, = \, V \left( x_1,\, x_2 \right)
\, + \, 4 \, \pi^2 \, x_{12}^{\ 2} \,
:\! \phi \left( x_1 \right) \phi \left( x_2 \right)
\! :
\eeq
\textit{allows to determine the
Taylor coefficients in} $x_1$ \textit{at}
$x_1=x_2$ \textit{of the
two terms in the right hand
side separately.}
\vspace{0.2in}

\textit{Sketch of proof.}\footnote{Complete proofs
of Propositions 2.2 and 2.3 will be published elsewhere.}
The (pseudo)harmonicity of $V$ (\ref{r2.7})
implies
\beq\label{r2.9}
\left( y\, \di_1 \right)^n
W \left( x_1,\, x \right)
\left| \,
\mathop{}\limits_{x_1 \, = \, x} \right.
\! = \, \left( y \, \di_1 \right)^n
V \left( x_1,\, x \right)
\left| \,
\mathop{}\limits_{x_1 \, = \, x} \right.
\! + y^2 \, n \left( n-1 \right) 4 \, \pi^2
:\! \left[
\left( y \, \di_x \right)^{n-2} \phi \left( x \right)
\right] \phi \left( x \right) \! :
\, . \
\eeq
In view of (\ref{r2.7})
$\Box_y \left( y \, \di_1 \right)^n$
$V \left( x_1,\, x \right)$
$\left| \, \mathop{}\limits_{x_1 \, = \, x} \right.$
$=\, 0\,$;
thus $\left( y \, \di_1 \right)^n$
$V \left( x_1,\, x \right)$
$\left| \, \mathop{}\limits_{x_1 \, = \, x} \right.$
appears as the harmonic part of the left hand
side of (\ref{r2.9}) viewed as a polynomial in $y$ and hence
is uniquely determined by
$W \left( x+y,\, x \right)\,$.$\Box\quad$
\vspace{0.2in}

It is clear from (\ref{r2.5}) that
$V \left( x_1,\, x_2 \right)$ is nonsingular for
coinciding arguments.
We can thus define a second
local field
\beq\label{r2.10}
\phi_2 \left( x \right) \, = \,
\frac{1}{2} \ V \left( x,\, x \right)
\, , \qquad
\eeq
of dimension $2\,$;
it
can be  a multiple of $\phi \left( x \right)$ only
if the ratios (\ref{1.8}) coincide.
Indeed, it follows from (\ref{1.7}) and (\ref{r2.5})
that
\[
\lvac \phi \left( x_1 \right) \phi \left( x_2 \right)
\rvac  = \la 12 \ra
\, , \
\lvac \phi_2 \left( x_1 \right) \phi \left( x_2 \right)
\rvac = \frac{c_3}{c_2} \, \la 12 \ra
\, , \
\lvac \phi_2 \left( x_1 \right) \phi_2 \left( x_2 \right)
\rvac = \frac{c_4}{c_2} \, \la 12 \ra
\, ; \qquad
\]
thus
\beq\label{r2.11}
\phi_2 \left( x \right) \, = \,
\lambda \, \phi \left( x \right)
\, (\, = \, \frac{c_3}{c_2} \phi \left( x \right) \, )
\quad \mathrm{implies} \quad
c_2 \, c_4 \, = \, c_3^2
\, . \qquad
\eeq

The preceding discussion admits an extension
to the $n$--point truncated function.
If we set, generalizing (\ref{r2.2}),
\beqa\label{r2.12}
&&
{\cal W}_{\, n}^{\, t} \lb x_1,\, ...,\, x_n \rb \, = \,
\frac{c_n}{2} \ \!\!\!
\Su_{\ \, \sigma \, \in \,
\mathrm{Perm} \LB 2 \, \dots \, n \RB } \!
\lb 1 \sigma_2 \rb
\, \BourWick{
\^1\sigma_2 \^1\sigma_3} \,
\dots
\, \BourWick{
\^1\sigma_{n-1} \^1\sigma_n} \,
\lb 1 \sigma_n \rb
\ , \quad
\nn &&
\BourWick{
\^1\sigma_i \^1\sigma_j}
\, = \,
\LB \!\!
\begin{array}{cc}
\lb \sigma_i \sigma_j \rb \quad \mathrm{for} \quad
\sigma_i \, < \, \sigma_j \\
\lb \sigma_j \sigma_i \rb \quad \mathrm{for} \quad
\sigma_j \, < \, \sigma_i
\end{array} \right.
\ , \qquad
n \, = \, 2,\, 3,\, 4,\, ...
\ . \qquad
\eeqa
then the field $\phi \left( x \right)$ of dimension
$2$ is unique if $c_n = c \, \alpha^n$ for some
$\alpha > 0\,$, $n$ $=\, 2,\, 3,\, ...\,$.

If we define
$V_1$ as a linear combination of normal products
of free (massless) fields,
\beq\label{r2.13}
V_1 \left( x_1,\, x_2 \right) \, = \,
\Su_{i \, = \, 1}^N \,
\alpha_i \, : \!
\varphi_i \left( x_1 \right) \varphi_i \left( x_2 \right)
\! :
\, , \quad
\eeq
and set \(\phi \left( x \right) = \phi_1 \left( x \right)
=\frac{1}{2} \ V_1 \left( x,\, x \right)\,\), then
we can reproduce (\ref{r2.12}) with
\beq\label{r2.14}
c_n \, = \, \Su_{i \, = \, 1}^N
\, \alpha_i^n
\, . \qquad
\eeq
Furthermore, we can introduce inductively
a series of bilocal and local fields
$V_n \left( x_1,\, x_2 \right)$ and
$\phi_n \left( x \right)$ of dimension
$\left( 1,\, 1 \right)$ and $2$ setting
\beqa\label{r2.15}
V_n \left( x_1,\, x_2 \right) \, = && \!\!\!\!\!\!\!
\mathop{\lim}\limits_{x_{34} \, \to \, 0} \,
\LB
4 \, \pi^2 x_{34}^{\ 2} \,
\Lb \vspe \!
V_1 \left( x_1,\, x_3 \right) \,
V_{n - 1} \left( x_2,\, x_4 \right)
\, - \,
c_n \, \lb \,
\lb 12 \rb \lb 34 \rb + \lb 14 \rb \lb 32 \rb \, \rb
\! \vspe \Rb
\RB \, =
\nn = && \!\!\!\!\!\!\!
\Su_{i \, = \, 1}^N \,
\alpha_i^n \, : \!
\varphi_i \left( x_1 \right) \varphi_i \left( x_2 \right)
\! :
\, , \quad
\phi_n \left( x \right) \, = \, \frac{1}{2}\
V_n \left( x,\, x \right)
\, . \qquad
\eeqa
Note
that the limit (\ref{r2.15}) is independent of the point
$x_3=x_4$ and
that the field $V$ appearing in the OPE (\ref{r2.3})
coincides with $V_2\,$.

The dimension of the space of different $d=2$ fields
$\phi_k \left( x \right)$ is equal to the number of
different values of $\alpha_i$ in (\ref{r2.13}).
To see this we note that the Gram determinant of
inner products
\beq\label{r2.16}
\lvac \phi_j \left( x_1 \right)
\phi_k \left( x_2 \right) \rvac
\, = \,
\frac{1}{2} \ \lb 12 \rb^2 \Su_{i \, = \, 1}^N \,
\alpha_i^{j+k}
\eeq
is a multiple of $\prod_{i \, = \,1}^N$
\, $\alpha_i^2$ \,
$\prod_{1 \, \leq \, j \, < \, k \, \leq \, N}$ \,
$\left( \alpha_j - \alpha_k \right)^2\,$.
\vspace{0.2in}

\textit{Remark 2.1}
Fields of type (\ref{r2.13})
have been studied in a different context (for bounded
$2$--dimensional fields) in \cite{Ba 99} \cite{GR 00}
where also infinite sums are admitted.
We restrict our discussion to finite $N$
since only in this case a stress energy tensor exists{\,}--
and is given by (\ref{n1.14}).
\vspace{0.2in}

From now on we shall restrict our discussion
to the simplest case of a single field $\phi$
of dimension $2$ and set
\beq\label{r2.17}
c_n \, = \, c \quad \mathrm{for} \quad
n \, = \, 2,\, 3,\, 4,\, ...
\eeq
(absorbing the possible factor $\alpha^n$
in the normalization of $\phi\,$).

The general form (\ref{r2.12}) of the truncated
$n$--point function can
in fact, be deduced. 
\vspace{0.2in}

\textbf{Proposition 2.3}
\textit{Let} $\phi \left( x \right)$ \textit{be a GCI
Wightman field of dimension} $2$ \textit{whose truncated}
$n$--\textit{point function is given by}
(\ref{r2.12}) \textit{with} $c_n = c$
\textit{for} $n\leq 4\,$.
\textit{Then
the limit}
\beq\label{+2.18}
V \left( x_1,\, x_2 \right)
\, = \,
\mathop{\lim}\limits_{\mathop{}\limits^{
\rho_{13} \, \to \, 0}_{\rho_{23} \, \to \, 0}} \,
\left( 2 \, \pi \right)^4 \, \rho_{13} \, \rho_{23} \,
\LB\vspe\!\vspe\!\right.
\phi \left( x_1 \right) \,
\phi \left( x_2 \right) \,
\phi \left( x_3 \right)
-
\la 13 \ra \phi \left( x_2 \right)
-
\la 23 \ra \phi \left( x_1 \right)
-
\la 123 \ra
\left.\vspe\!\!\RB
\,
\eeq
\textit{exists, does not depend on} $x_3\,$,
\textit{and defines a harmonic in each argument
bilocal field} $V \left( x_1,\, x_2 \right)\,$.
\textit{Furthermore,
the truncated} $n$--\textit{point
functions of} $\phi$ \textit{will be given by}
(\ref{r2.12}) \textit{for all} $n\,$.
\vspace{0.2in}

\textit{Sketch of proof.}
Equation (\ref{r2.2}) and the conservation of the
stress-energy tensor (see Sec. 3) imply that (\ref{r2.12}) is valid for
$n \leq 6$.
The $1$--loop expression for the $6$--point function allows
to derive (\ref{+2.18}).
The expression for the correlation function (\ref{r2.5}) of two $V$'s
satisfies the d'Alembert equation in each
argument.
By virtue of Proposition 2.1
the operator field $V$ obeys this equation
in its entire domain.
Eq. (\ref{r2.12}) for $n\leq 6$ implies an
expansion of the form
\beqa\label{++2.18}
&& \!\!\!\!\!\!\!
\phi \left( x_1 \right)
\phi \left( x_2 \right)
\phi \left( x_3 \right)
\, \rvac
\, = \,
\la 123 \ra \, \rvac
+
\nn && \!\!\!\!\!\!\!
\qquad
\ +
\Su_{\mathop{}\limits^{i \, = \, 1,\, 2,\, 3}_{
j \, < \, k \, , \ j \neq i \neq k}} \!
\left\{ \!\vspe
\la jk \ra \phi \left( x_i \right)
+ \,
\BourWick{
\^1i \^1j} \ \,
\BourWick{
\^1i \^1k} \ \,
V \left( x_j,\, x_k \right)
+
\left( jk \right) :\! V \left( x_j,\, x_k \right)
\phi \left( x_i \right)\! :
\!\vspe \right\} \rvac
+
\nn && \!\!\!\!\!\!\!
\qquad
\ +
:\! \phi \left( x_1 \right)
\phi \left( x_2 \right)
\phi \left( x_3 \right) \! :
\, \rvac
\,  \qquad
\eeqa
($\left( i,\, j,\, k \right)$
form permutations
of $1,\, 2,\, 3\,$).
The result then follows.$\quad\Box$
\vspace{0.2in}

\textit{Remark 2.2}
If we drop the requirement of Wightman positivity - which implies the
validity of the stress-energy tensor conservation as an operator
equation - then the general form of the truncated 5-point function would be
\beq\label{r2.19}
\W_5^{\, t} \left( x_1,\, ...,\, x_5 \right)
\, = \,
\lambda \,
\W_5^{\, t} \left( \mathrm{\ref{r2.12}} \right)
+
4 \, \pi^2 \, c \, \left( 1 - \lambda \right)
\Su_{1 \, \leq \, i \, < \, j \, \leq \, 5} \,
\rho_{ij} \,
\mathop{\prod}\limits_{
\mathop{}\limits_{1 \, \leq \, k \, \leq \, 5}^{
j \, \neq \, k \, \neq \, j}} \,
\BourWick{
\^1i \, \^1k} \
\BourWick{
\^1k \, \^1j}
\, , \quad
\lambda \, \in \, \R
\, . \qquad
\eeq
We note that the $1$--dimensional time--like
restriction $\phi\left( t,\, \mathbf{0} \right)$
of $\phi \left( x \right)$ satisfies all properties
of the chiral stress energy tensor in a $2$D CFT.
It follows that all restricted truncated functions
should have the form (\ref{r2.12}).
This is satisfied by (\ref{r2.19})
(for our choice of constants) because of a non--trivial
identity between the two terms in the $1$--dimensional
case.
\vspace{0.2in}

\textbf{Corollary 2.4}
\textit{Under the assumptions of Proposition 2.3
one can prove (using also Proposition 2.2) that
the field algebra of} $\phi \left( x \right)$
\textit{coincides with the algebra of the bilocal field}
$V \left( x_1,\, x_2 \right)\,$.
\vspace{0.2in}

Demanding that the truncated $n$--point function of $\phi$
for $n \geq 3$ is strictly less singular in $x_{ij}$
than its $2$--point function we have taken into
account a necessary condition for Wightman positivity.
We shall
prove a necessary and sufficient condition for
positivity in Sec. 5.
\vspace{0.2in}

{\it Remark 2.3}
If we rescale the field $\phi$ by a factor
$c^{-\frac{1}{2}}$ and let $c \to \infty$ we recover
the case of a generalized free field of dimension $2\,$:
\beqa\label{r2.20}
&& \!\!\!\!\!\!\!
\mathrm{if} \quad \widehat{\phi} \lb x \rb
= \frac{1}{\sqrt{c}} \phi \lb x \rb \quad \mathrm{then}
\quad
\nn && \!\!\!\!\!\!\!
\mathop{\lim} \limits_{c \to \infty} \,
\la 0 \l| \,
\widehat{\phi} \lb x_1 \rb \widehat{\phi} \lb x_2 \rb
\widehat{\phi} \lb x_3 \rb \widehat{\phi} \lb x_4 \rb
\ \r| 0 \ra
\, = \,
\la 12 \ra_1 \la 34 \ra_1 \, + \,
\la 13 \ra_1 \la 24 \ra_1 \, + \,
\la 14 \ra_1 \la 23 \ra_1
\ , \qquad \quad
\eeqa
where $\la ij \ra_1 = \frac{1}{2} \, \lb ij \rb^2\,$.




\sectionnew{
Expansion of $V \left( x_1,\, x_2 \right)$ in
local fields.
Infinite set of conserved tensor currents}

We shall now demonstrate that our model possesses an
infinite  number of conserved local tensor currents.
More precisely, the bilocal field $V \lb x_1,\, x_2 \rb$
can be expanded in a series of even rank, conserved
symmetric traceless tensor fields
$T_{2l} \lb x,\, \zeta \rb$ (\ref{1.12}) (of twist
$=\ $dimension $-\ $rank $=\ 2\,$):
\beq\label{2.13}
V \lb x_1,\, x_2 \rb \, = \,
2 \, \Su_{l \, = \, 0}^{\infty} \,
C_l \, K_l \lb x_{12} \! \cdot \! \partial_2 \, ,\,
\rho_{12} \, \Box_2 \rb
T_{2l} \lb x_2,\, x_{12} \rb
\ , \qquad
\eeq
reproducing the $4$--point function (\ref{r2.5}).
Here
\beq\label{2.14}
K_l \lb s,\, t \rb \, = \,
\frac{\lb 2l + 1 \rb !}{\lb l! \rb^{\, 2}} \,
\mathop{\int}\limits_{\!\!\!\!\!\! 0}^{\;\;\ 1}
\mathrm{d}\alpha \ \alpha^l \lb 1 - \alpha \rb^l
e^{\alpha s} \
\Su_{n \, = \, 0}^{\infty} \
\frac{\lb - \, \frac{\alpha \, \lb 1 - \alpha \rb}{4}
\, t \rb^n}{n! \, \lb 2l +1 \rb_n }
\ , \quad
(\, K_l \lb 0,\, 0 \rb \, = \, 1 \, )
\ , \quad
\qquad
\eeq
$\partial_2$ is the derivative in $x_2$ for  fixed
$x_{12}\,$, $\Box_2$ is the corresponding d'Alembert
operator,
\( \left( \nu \right)_n \, = \,
\frac{\Gamma \left( n+\nu \right)}{
\Gamma \left( \nu \right)}\,\);
it is chosen to transform the $2$--point
function
$\la 0 \l| T_{2l} \lb x_2,\, \zeta_2 \rb
T_{2l} \lb x_3,\, \zeta_3 \rb \r| 0 \ra$
into a $3$--point function:
\beq\label{2.15}
K_l \lb x_{12} \! \cdot \! \partial_2 \, ,\,
\rho_{12} \, \Box_2 \rb \,
\frac{\lb x_{12} \! \cdot \! r \lb x_{23} \rb
\! \cdot \! \zeta \rb^{2l}}{\rho_{23}^{2l+2}}
\, = \, \frac{\lb X \!\! \cdot \! \zeta \rb^{2l}}{\rho_{13} \,
\rho_{23}}
\ , \qquad
\eeq
where
\beq\label{2.16}
\xi \cdot r \lb x_{23} \rb \cdot \zeta \, = \,
\xi \cdot \zeta \, - \, 2 \
\frac{\lb \xi \! \cdot \!  x_{23} \rb \lb \zeta \! \cdot \! x_{23}
\rb}{\rho_{23}}
\ , \quad
X \, := \, X^3_{12} \, := \,
\frac{x_{13}}{\rho_{13}} \, - \,
\frac{x_{23}}{\rho_{23}}
\ , \quad
(\, X^2 \, = \,
\frac{\rho_{12}}{\rho_{13} \, \rho_{23}}  \, )
\ . \qquad
\eeq
In verifying (\ref{2.15}) (see \cite{DO 01}) one applies
the relation
\beqa\label{no1}
\lb \frac{\Box_y}{4} \rb^n \frac{\lb y \! \cdot \! \zeta \rb^m}{
\lb y^2 \rb^{\nu}}
\, = \, \frac{\lb \nu \rb_n \lb \nu - m - 1 \rb_n}{
\lb y^2 \rb^{n + \nu}}
\, \lb y \! \cdot \! \zeta \rb^m
\quad \mathrm{for} \quad
\zeta^2 \ = \ 0
\qquad
\nonumber
\eeqa
(used for $y \ = \ x_{23} \ + \ \alpha \, x_{12}\,$).
In order to compute individual contribution of $T_{2l}$
to the $4$--point function of $\phi$ we need the
$3$--point function
\beq\label{2.17}
\la 0 \l| \phi \lb x_1 \rb \phi \lb x_2 \rb
T_{2l} \lb x_3,\, \zeta \rb \r| 0 \ra \, = \,
N_l \ C_l \, \la 12 \ra
\lb X^2 \rb^{l+1} \lb \zeta^2 \rb^l
C_{2l}^1 \! \lb \!
\widehat{X} \! \cdot \! \widehat{\zeta}
\rb
\, , \quad
\widehat{X} \, := \,
\frac{X}{\,\sqrt{X^2}\,}
\ , \qquad
\eeq
where $N_l > 0\,$, $C_n^1 \lb z \rb$ is the Gegenbauer
polynomial satisfying
\beq\label{2.18}
\LB
\lb 1 - z^2 \rb \frac{\mathrm{d}^2}{\mathrm{d}z^2}
\, - \, 3 z \, \frac{\mathrm{d}}{\mathrm{d}z} \, + \,
n \lb n + 2 \rb
\RB \, C_n^1 \lb z \rb \, = \, 0
\ , \quad C_n^1 \lb 1 \rb \, = \, n \, + \, 1
\ . \qquad
\eeq
Writing the normalization constant in (\ref{2.17}) as
a product, $N_l C_l\,$, we exploit the fact that the
$3$--point function vanishes whenever the structure
constant $C_l = 0\,$.
\vspace{0.2in}

{\it Remark 3.1}
The  homogeneous polynomial
$H_{2l} \lb x,\, \zeta \rb$
$= \lb x^2 \, \zeta^2 \rb^l$
$C_{2l}^1 \! \lb \!
\widehat{x} \! \cdot \! \widehat{\zeta}
\rb$
is the harmonic extension of the monomial
$\lb 2 \, x \! \cdot \! \zeta \rb^{2l}$ defined
on the light cone $\zeta^2 = 0$ (cf. \cite{BT77}):
\beqa\label{2.19}
&& \!\!\!\!\!
\Box_{\zeta} H_{2l} \lb x,\, \zeta \rb \, = \,
\lb x^2 \rb^l \lb \zeta^2 \rb^{l-1} \, \times
\nn && \!\!\!\!\! \qquad\qquad\qquad\quad\,
\times \,
\LB
\lb 1 - z^2 \rb \frac{\mathrm{d}^2}{\mathrm{d}z^2}
\, C_{2l}^1 \lb z \rb
\, - \, 3 z \, \frac{\mathrm{d}}{\mathrm{d}z}
\, C_{2l}^1 \lb z \rb  \, + \,
4l \lb l + 1 \rb \, C_{2l}^1 \lb z \rb
\RB \, = \,
0
\quad
\nn && \!\!\!\!\!
(\, \mathrm{for} \ z \, = \,
\widehat{x} \! \cdot \! \widehat{\zeta}
\, ) \ , \quad
\left. H_{2l} \lb x,\, \zeta \rb
\mathop{}\limits_{} \! \right|
\! \mathop{}\limits_{\zeta^2 \, = \, 0}
\, = \, \lb 2 \, x \! \cdot \! \zeta \rb^{2l}
\ . \qquad
\eeqa
Similarly, the $2$--point function
$\la 0 \l| T_{2l} \lb x_1,\, \zeta_1 \rb
T_{2l} \lb x_2,\, \zeta_2 \rb \r| 0 \ra$
is proportional to
$\rho_{12}^{-2l-2}$
$\lb \zeta_1^{\, 2} \zeta_2^{\, 2} \rb^l$
$C_{2l}^1 \lb \widehat{\zeta}_1 \! \cdot \! \widehat{\zeta}_2 -
2 \frac{\lb \widehat{\zeta}_1 \cdot \, x_{12} \rb
\lb \widehat{\zeta}_2 \cdot \, x_{12} \rb}{\rho_{12}} \rb \,$.
\vspace{0.2in}

Inserting (\ref{r2.3}) in the $4$--point function
(\ref{r2.1}) (\ref{r2.2}) and using (\ref{r2.5})
and the expansion (\ref{2.13}) for
$V \lb x_3,\, x_4 \rb$
we find
\beqa\label{2.20}
&& \!\!\!\!\!\!\!\!\!\!
\la 0 \l| \,
\phi \lb x_1 \rb \phi \lb x_2 \rb
V \lb x_3,\, x_4 \rb \r| 0 \ra
\, = \,  c\, \lb 12 \rb
\lb \lb 13 \rb \lb 24 \rb \, + \, \lb 14 \rb \lb 23 \rb \rb
\, = \,
\nn && \!\!\!\!\!\!\!\!\!\! \qquad \
= \,
2 \, \Su_{l \, = \, 0}^{\infty} \,
C_l \, K_l \lb x_{34} \! \cdot \! \partial_4 \, ,\,
\rho_{34} \, \Box_4 \rb
\la 0 \l| \,
\phi \lb x_1 \rb \phi \lb x_2 \rb
T_{2l} \lb x_4,\, x_{34} \rb \r| 0 \ra \, = \,
\nn && \!\!\!\!\!\!\!\!\!\! \qquad \
= \,
4 \la  12 \ra \,
\Su_{l \, = \, 0}^{\infty} \, N_l \, C_l^2 \
\frac{\lb 4l + 1 \rb !}{\lb 2l \rb !^2}
\mathop{\int}\limits_{\!\!\!\!\!\! 0}^{\;\;\ 1}
\mathrm{d}\alpha \ \alpha^{2l} \lb 1 - \alpha \rb^{2l} \
\frac{\lb - \, \frac{\alpha \, \lb 1 - \alpha \rb}{4}
\, \rho_{34} \, \Box_4 \rb^n}{n! \, \lb 2l +1 \rb_n }
\, \times \,
\nn && \!\!\!\!\!\!\!\!\!\! \qquad \
\qquad \qquad \qquad \qquad
\times \ \rho_{34}^{\, l} \ \lb X_y^2 \rb^{l + 1} \
C_{2l}^1 \! \lb
\widehat{X}_y \! \cdot \! \widehat{x}_{34} \rb
\ , \qquad
\nn && \!\!\!\!\!\!\!\!\!\!
X_y \, = \, \frac{x_1 - y}{\rho_{1y}}
\, - \, \frac{x_2 - y}{\rho_{2y}}
\ , \quad
y \, = \, x_4  \, + \, \alpha \, x_{34}
\ , \quad
\nn && \!\!\!\!\!\!\!\!\!\!
\rho_{iy} \, = \, \rho_{i4} \lb 1 - \alpha \rb
\, + \, \alpha \rho_{i3} \, - \,
\alpha \lb 1 - \alpha \rb \rho_{34}
\ , \quad i \, = \, 1,\, 2
\ . \qquad
\eeqa

It will be convenient for what follows to substitute the second
conformally invariant
cross ratio $\eta_2$ (\ref{1.4}) by the
difference $\epsilon = 1 - \eta_2$ which tends to zero
for $x_{34} \to 0$ (or $x_{12} \to 0\,$):
\beq\label{2.21}
\epsilon \, = \, 1 \, - \, \eta_2 \, ( \, = \,
O \lb x_{34} \rb \, = \, O \lb x_{12} \rb \, )
\ . \qquad
\eeq
\vspace{0.2in}

{\bf Proposition 3.1}
\textit{For}
\beq\label{2.22}
N_l \, C_l^2 \, = \,
\lb \!\!\!
\begin{array} {l} 4 l \\ 2 l \end{array}
\!\! \rb^{\! -1}
\qquad
\eeq
\textit{the contribution of} $V \lb x_3,\, x_4 \rb$
\textit{to the} $4$--\textit{point function} (\ref{r2.1})
\textit{is reproduced by the superposition} (\ref{2.20})
\textit{of} $3$--\textit{point functions of the twist} $2$
\textit{fields} $T_{2l}\,$
\beqa\label{2.23}
\frac{\la 0 \l| V \lb x_1,\, x_2 \rb
V \lb x_3,\, x_4 \rb  \r| 0 \ra}{\lb 13 \rb \lb 24 \rb}
\,
&& \!\!\!\!\!\!\!\!
= \,
c \, \lb 1 \, + \, \frac{1}{1 \! - \! \epsilon} \rb
\, = \,
\nn && \!\!\!\!\!\!\!\!
= \,
2 c \, \Su_{l \, = \, 0}^{\infty} \, \lb 4 l + 1 \rb
\mathop{\int}\limits_{\!\!\!\!\!\! 0}^{\;\;\ 1}
\Lb \frac{\epsilon \, \alpha \, \lb 1 - \alpha \rb}{
1 - \epsilon \, \alpha} \Rb^{2l} \!
\frac{\mathrm{d}\alpha}{1 - \epsilon \, \alpha}
\ . \qquad \quad
\eeqa
\vspace{0.2in}

The \textit{proof} of this statement is given in
Appendix A.

The Ward--Takahashi identity for the time--ordered
$3$--point function of the stress--energy tensor allows
to compute the normalization $N_1 C_1$ of the Wightman
function (\ref{2.17}):
\beq\label{2.24}
\la 0 \l| \phi \lb x_1 \rb \phi \lb x_2 \rb
T_2 \lb x_3,\, \zeta \rb \, \r| 0 \ra \, = \,
\frac{\la 12 \ra}{3 \, \pi^2} \ X^2
\lb X^2 \zeta^2  \, - \,
4 \lb X \!\! \cdot \! \zeta \rb^2 \rb
\ , \quad \mathrm{i.e.} \quad
N_1 \, C_1 \, = \, \frac{-1}{3 \, \pi^2}
\ . \qquad
\eeq
Comparing with (\ref{2.22}) we find that both $N_1$ and
$C_1$ are transcendental, only the combination $N_1 C_1^2$
being rational:
\beq\label{2.25}
C_1 \, = \, - \ \frac{\pi^2}{2} \ , \quad
N_1 \, = \, \frac{2}{3 \, \pi^4} \ , \quad
(\, N_1 \, C_1^2 \, = \, \frac{1}{6} \, )
\ . \qquad
\eeq
\vspace{0.2in}

\textit{Remark 3.2}
It is instructive to note that the contribution
of each $T_{2l}$ to the ratio (\ref{2.23})
(given by the $l$th term in the right hand side)
involves a logarithmic function in $1-\epsilon$
(see Appendix A) while the infinite sum is a rational
function of $\epsilon\,$.




\sectionnew{
The infinite dimensional Lie algebra of field modes
and its bilocal realization}

The conformal compactification
\( \M = \left. \Sr^3 \! \times \! \Sr^1 \right/ \Z_2 \)
of Minkowski space $M = \R^{3,1}$ gives rise to a natural
notion of conformal energy, the generator of (isometric)
rotation of the time--like circle $\Sr^1\,$, and of an
associated discrete basis of field modes.
We shall parametrize $\M$ following (\cite{Tod86}) in terms
of complex coordinates
$z = \lb z_a,\, a = 1,\, 2,\, 3,\, 4 \rb$
fixed by the involution
$z\mapsto z^*:=\frac{\,\overline{z}\,}{\overline{z}^{\, 2}}\,$:
\beq\label{3.1}
\overline{M} \, = \,
\LB z \, = \,
\lb z_a \, \in \, \C,\, a \, = \, 1,\, ...,\, 4 \rb
\ ; \quad
z_a^* \, := \, \frac{\overline{z}_a}{\overline{z}^{\, 2}}
\, = \, z_a
\quad (\, z^2 \, = \, \Su_{a} \, z_a^2 \, =: \,
\mathbf{z}^2 \, + \, z_4^2 \, ) \ \RB
\ . \qquad
\eeq
This condition implies the property
\beq\label{3.2}
z^2 \, \overline{z}^{\, 2} \, = \, 1 \ , \quad
\frac{z_a \, z_b}{z^2} \, = \,
\overline{z}_a \, z_b \, = \,
z_a \, \overline{z}_b \, \in \R \quad
\mathrm{for} \quad z \, \in \, \M
\qquad
\eeq
which, in turn, characterizes this parametrization of
$\overline{M}\,$.
We choose the embedding map $M \subset \overline{M}$ as
\beq\label{3.3}
M \, \ni \, \lb x^0,\, \mathbf{x} \rb \, \mapsto \,
\mathbf{z} \, = \, \omega^{-1} \lb x \rb \, \mathbf{x}
\ , \quad z_4 \, = \,
\frac{1 \! - \! x^2}{2 \, \omega \lb x \rb} \ , \quad
\omega \lb x \rb \, = \, \frac{1 \! + \! x^2}{2} \, - \,
i \, x^0
\ . \qquad
\eeq
Clearly, $z$ defined by (\ref{3.3}) satisfies (\ref{3.2});
in particular,
\beq\label{3.4}
z^2 \, = \,
\frac{\ \overline{\omega \lb x \rb \!}\ }{\omega \lb x \rb}
\, = \,
\frac{\lb 1 \! + \! i \, x^0 \rb^2 + \mathbf{x}^2}{
\lb 1 \! - \! i \, x^0 \rb^2 + \mathbf{x}^2} \, = \,
\frac{1}{\overline{z}^{\, 2}} \ , \quad
\l| \! z^2 \r| \, = \, z \! \cdot \! \overline{z} \, = \, 1
\quad (\, \mathrm{for} \quad z \, \in \, \M \, )
\ . \qquad
\eeq
In order to write down the inverse transformation it is
convenient to present $z$ in terms of a complex quaternion
(or, equivalently, an element of
$\mathrm{U} \lb 2 \rb\,$-- see \cite{Uhl 63}):
\beq\label{3.5}
q \, z \, = \, z_4 \, + \, \mathbf{z} \, \mathbf{q} \ , \quad
q_i \, q_j \, = \, \epsilon_{ijk} \, q_k
\, - \, \delta_{ij} \quad
\quad (\, \mathrm{i.e.} \quad q_1 \, q_2 \, = \,
- \, q_2 \, q_1 \, = \, q_3 \, , \quad \mathrm{etc.} \, )
\ . \qquad
\eeq
The cone at infinity,
$K_{\infty} = \left. \overline{M} \, \right\backslash M\,$,
consists of the quaternions $q z \in \overline{M}$
for which $1 + q z$ is not invertible
\beq\label{3.6}
q \, z \, \in \, K_{\infty} \quad \mathrm{iff} \quad
2 \, \omega^{-1}_z \, := \, \lb 1 + q \, z \rb
\lb 1 + q^+ \, z \rb \, = \, \lb 1 + z_4 \rb^2 \, + \,
\mathbf{z}^2 \, = \, 0 \quad
(\, q^+ \, z \, = \, z_4 \, - \, \mathbf{q} \, \mathbf{z} \, )
\ . \qquad
\eeq
For $q z \notin K_{\infty}$ we can set
\beq\label{3.7}
i \, \widetilde{x} \, := \,
i \, x_0 \, + \, \mathbf{q} \, \mathbf{x} \, = \,
\frac{q \, z - 1}{q \, z + 1} \quad \mathrm{or} \quad
i \, x_0 \, = \, \omega_z \, \frac{z^2 \! - \! 1}{2}
\ , \quad \mathbf{x} \, = \ \omega_z \, \mathbf{z} \, = \,
\frac{2 \, \mathbf{z}}{\lb 1 \! + \! z_4 \rb^2 +
\mathbf{z}^2}
\ . \qquad
\eeq
We shall use the fact
that the flat metric on $\overline{M}$ is
related to the Poincar\'{e} invariant metric on $M$ by the
complex conformal factor $\omega$ (\ref{3.3})
\beq\label{3.8}
\mathrm{d}z^2 \, = \, \mathbf{\mathrm{d}z}^2 + \,
\mathrm{d} z_4^{\, 2} \, = \, \omega^{-2} \! \lb x \rb \,
\mathrm{d} x^2 \quad
(\, \mathrm{d}x^2 \, = \, \mathbf{\mathrm{d}x}^2 - \,
\mathrm{d}x_0^{\, 2} \, )
\ . \qquad
\eeq
To a scalar field $\phi_M \lb x \rb$ of dimension $d$ in
Minkowski space we make correspond an \textit{analytic}
$z$--\textit{picture field} $\phi \lb z \rb$ defined by:
\beq\label{3.9}
\phi \lb z \rb \, = \, \lb 2 \, \pi \rb^d \, \omega_z^d \
\phi_M \! \lb x \lb z \rb \rb
\quad (\, \omega_z \, = \,
\frac{2}{\lb 1 \! + \! z_4 \rb^2 + \mathbf{z}^2} \, = \,
\omega \lb x \lb z \rb \rb \, )
\qquad
\eeq
for $x \lb z \rb$ given by (\ref{3.7}).
The term \textit{analytic} is justified by the fact that
energy positivity implies analyticity of the vector valued
function $\left. \left. \phi \lb z \rb \r| 0 \ra$
for ${\l| \! z \r| \, }^2 < 1\,$.
Indeed, the future tube
$T_+ $ $= \LB \, \zeta \in \C^4 \, ; \
\mathrm{Im} \, \zeta^0
> \l| \! \mathrm{Im} \, \underline{\zeta} \, \r| \, \RB\,$,
the analyticity domain of
$\left. \left. \phi_M \lb \zeta \rb \r| 0 \ra$
(see \cite{SW64-00}), is mapped into a complex
neighbourhood  $\SMB_+$ of the $4$--dimensional unit ball
$\BB^4\,$;
more precisely, we have
\beqa\label{3.10}
&&
\BB^4 \, = \, \LB \,
\xi \, \in \, \R^4 \ ; \quad
\xi^2 \, := \, \underline{\xi}^2 \, + \, \xi_4^2
\, < \, 1 \, \RB \ , \quad
\nn &&
\BB^4 \! \times \Sr^1 \! \left/ \, \Z_2 \right.
\, = \, \LB \,
z \, = \, \xi \, e^{i \tau} \ ; \quad
\xi \, \in \, \BB^4 ,\, \tau \, \in \, \R \,
\RB \, \subset \, \SMB_+
\ . \qquad
\eeqa
Note that $\overline{M}$ appears as the boundary of the
$5$--dimensional manifold
$\BB^4 \! \times \Sr^1 \! \left/ \, \Z^2 \right. \,$:
\beq\label{3.11}
z \, \in \, \M \quad \mathrm{iff} \quad
z \, = \, e^{i \tau} \, \widehat{z} \ , \quad
\tau \, \in \, \R
\ , \quad
\widehat{z} \, \in \, \Sr^3 \, = \,
\LB \, \widehat{z} \, \in \, \R^4 \ ; \quad
\widehat{z}^{\, 2} \, = \, 1 \, \RB
\ . \qquad
\eeq
The \textit{conformal Hamiltonian} $H$ is, in this picture,
nothing but the (hermitian) generator of translation in
$\tau \,$:
\beq\label{3.12}
e^{i H t} \, \phi \lb z \rb \, e^{- \, i H t}
\, = \, e^{i t d} \, \phi \lb e^{i t} \, z \rb
\quad \mathrm{or} \quad
\Lb \, H , \phi \lb z \rb \, \Rb \, = \,
\lb d + z_a \, \frac{\partial}{\partial z_a} \rb
\phi \lb z \rb \ , \quad \left. \left. H \r| 0 \ra \, = \, 0
\ . \qquad
\eeq
The decomposition of $\phi$ into eigenmodes of $H$ reads
\beq\label{3.13}
\phi \lb z \rb \, = \, \Su_{n \, \in \, \Z}
\, \phi_n \lb z \rb \ , \quad
\Lb \, \phi_n \! \lb z \rb , H \, \Rb \, = \, n \,
\phi_n \! \lb z \rb
\ . \qquad
\eeq
The \textit{modes} $\phi_n \lb z \rb$ can be written as
power series in $z_a$ and$\frac{1}{z^2}$ that are
homogeneous in $z$ of degree $-n-d\,$.

For a free field $\varphi \lb z \rb$ of dimension $d=1$
the modes $\varphi_{\pm n}$ are homogeneous harmonic
polynomials spanning a space of dimension $n^2$ (as a
space of $\mathrm{SO} \lb 4 \rb$ symmetric traceless
tensors of rank $n-1\,$:
$\lb \!\!\! \begin{array}{c} n+2 \\ 3 \end{array} \!\!\! \rb$
$- \, \lb \!\!\! \begin{array}{c} n \\ 3 \end{array} \!\!\! \rb$
$= \, n^2\,$);  in particular,
$\varphi_0 \! \lb z \rb$ $= 0\,$,
$\varphi_1 \! \lb z \rb$ $= \frac{a_1}{z^2}\,$,
$\varphi_{-1} \! \lb z \rb$ $= a_{-1}\,$,
$\varphi_2 \! \lb z \rb$
$= \frac{a_2^{\mu} z_{\mu}}{\lb z^2 \rb^2}\,$,
$\varphi_{-2} \! \lb z \rb$ $= a_{-2}^{\mu} z_{\mu}$
etc.
They are subject to the canonical commutation relations
\cite{Tod86}
\beqa\label{3.14}
&& \!\!\!\!\!\!
\Lb \, \varphi_n \! \lb z \rb , \varphi_m \! \lb w \rb
\, \Rb \, = \,
\frac{\lb w^2 \rb^{\frac{n-1}{2}}}{
\lb z^2 \rb^{\frac{n+1}{2}}} \
C_{\left| n \right| -1}^1 \lb \widehat{z} \! \cdot \! \widehat{w} \rb \,
\epsilon \left( n \right) \, \delta_{n,-m} \ , \quad
(\, z \, = \, \sqrt{z^{2 \,}} \, \widehat{z} \,)
\nn && \!\!\!\!\!\!
\epsilon \left( n \right) \, = \,
\LB
\begin{array}{r}
1 \quad \mathrm{for} \quad n \, > \, 0 \\
0 \quad \mathrm{for} \quad n \, = \, 0 \\
-1 \quad \mathrm{for} \quad n \, < \, 0
\end{array}
\right.
\ . \qquad
\eeqa
(Here one uses the fact that the $2$--point function
$\la 0 \l| \varphi \lb z \rb \varphi \lb w \rb \r| 0 \ra$
$= \frac{1}{\lb z \! - \! w \rb^2}$ appears as a generating
function for the Gegenbauer polynomials defined in
(\ref{2.18}).)

One can expand
the bilocal field $V$
in modes
\(
V =
\Su_{\mathop{}\limits^{n,\, m}_{n \, \neq \, 0 \, \neq \, m}}
V_{nm}
\,
\),
which
behave as
products of $\varphi$--modes:
\beqa\label{r4.15}
&& \!\!\!\!\!\!\!
\Delta_z \,
V_{nm} \left( z,\, w \right) \, = \, 0 \, = \,
\Delta_w \,
V_{nm} \left( z,\, w \right)
\, , \quad
\nn && \!\!\!\!\!\!\!
\left( z \! \cdot \! \frac{\di}{\di z} \,
+ n + 1  \right)
V_{nm} \left( z,\, w \right)
\, = \, 0 \, = \,
\left( w \! \cdot \! \frac{\di}{\di w} \,
+ m + 1  \right)
V_{nm} \left( z,\, w \right)
\, . \qquad
\eeqa
(The homogeneity condition only agrees with the
Laplace equation if we set $V_{0m}$ $=\, 0$ $=\, V_{n0}\,$.)
The modes of the $d=2$ field $\phi$ are most conveniently
expressed as infinite sums of $V$--modes:
\beq\label{r4.16}
2 \, \phi_n \left( z \right) \, = \,
\Su_{\nu \, \in \, \Z} \,
V_{\nu,\, n-\nu} \left( z,\, z \right)
\quad
(\, V_{mn} \left( z,\, z \right)
\, = \,
V_{nm} \left( z,\, z \right) \, )
\qquad
\eeq
The components $V_{\nu,\, n-\nu} \left( z,\, z \right)$
of $\phi_n \left( z \right)$
(unlike those of $\varphi_n \left( z \right)$) span
an infinite dimensional space.
This is a common feature for scalar fields
of dimension $d > 1$
(more generally, for elementary conformal fields
of weight $\left( j_1,\, j_2;\, d \right)$
with $d$ $\geq\, j_1$ $+\, j_2$ $+\, 2\,$,
in the notation of \cite{Mack77b} and \cite{NT 01},
which, as a result, cannot obey a free field equation).
It is all the more remarkable that the state space for
a given energy $n$ is always finite dimensional.
This is a consequence
of the analyticity of the vector valued
function $V_{nm} \left( z,\, w \right) \rvac$
for $z,\, w$ $\in \, \SMB_+\,$.
Indeed, it then follows from (\ref{3.10})
and (\ref{r4.15}) that
\beq\label{r4.17}
V_{nm} \left( z,\, w \right) \rvac \, = \, 0
\quad \mathrm{if} \quad n \, \geq \, 0
\quad \mathrm{or} \quad m \, \geq \, 0
\, . \qquad
\eeq
Consequently, only
$\left( n-1 \right)$ terms of the
infinite sum (\ref{r4.16}) contribute
to the vector $\phi_{-n} \left( z \right)\rvac\,$:
$2 \, \phi_{-n}\left( z \right)\rvac$
$= \, \su_{\nu \, = \, 1}^{n-1}$
$V_{-\nu,\, \nu - n} \left( z,\, z \right)\rvac\,$.

In order to display the identity of the vacuum
state spaces of $\phi$ and $V\,$,
guaranteed by Corollary 2.4 we need
to include the composite twist $2$ fields
$T_{2l} \left( z,\, \zeta \right)$
in the operator algebra of $\phi\,$.
Here is the realization of the four lowest  energy
spaces in the two pictures.
Setting for the vacuum Hilbert space
\[
{\cal H} \, = \, {\cal H}_0
\, \oplus \,
\mathop{\bigoplus}\limits_{n \, = \, 2}^{\infty} \,
{\cal H}_n
\, , \quad
\left( H - n \right) \, {\cal H}_n \, = \, 0
\quad
(\, \dim \, {\cal H}_0 \, = \, \dim \, {\cal H}_2
\, = \, 1 \, )
\qquad
\]
we can write down a basis in ${\cal H}_2\,$, ${\cal H}_3$
and ${\cal H}_4$ as follows:
\beqa\label{no2}
&& \!\!\!\!\!\!\!
\phi_{-2} \, \rvac \, = \,
\frac{1}{2} \ V_{-1,\, -1} \, \rvac
\, ; \quad
\phi_{-3}^{\, a} \, z_a \, \rvac \, = \,
V_{-2,\, -1}^{\, a} \, z_a \, \rvac
\, (\, = \, z_a V_{-1,\, -2}^{\, a} \, \rvac \, )
\ ;
\nn && \!\!\!\!\!\!\!
\LB
\, \phi_{-4}^{\, ab} \, \rvac,\,
T_2 \left( 0,\, \zeta \right) \, \rvac,\,
\phi_{-2}^{\, 2} \, \rvac \,
\RB
\, \sim \,
\LB
\, V_{-2,\, -2}^{\, a b} \, z_a \, z_b \, \rvac,\,
V_{-3,\, -1}^{\, a b} \, z_a \, z_b \, \rvac,\,
V_{-1,\, -1}^{\, 2} \, \rvac
\,
\RB
\, . \qquad
\nonumber
\eeqa

The difficulty in describing the full state
space ${\cal H}$ in such a manner stems from the fact that
the modes of $\phi$ do not span an (infinite dimensional) Lie
algebra: the commutator
$\Lb \phi \lb z_1 \rb , \phi \lb z_2 \rb \Rb$
also involves all twist $2$ conserved tensors
$T_{2l} \lb z_2,\, z_{12} \rb$ (and their derivatives in the
first argument).
$T_{2l}$ $\lb l = 0,\, 1,\, ... \rb$ together with the unit
operator exhaust, in fact, the singular terms in the OPE
$\left. \left. \phi \lb z \rb \phi \lb w \rb \r| 0 \ra\,$.
The resulting commutator algebra simplifies drastically for
collinear $z_j = \zeta_j \, e$ ($e^2 = 1$): it then reduces
to the Virasoro algebra:
\beq\label{3.18}
\Lb \, L_n , L_m \, \Rb \, = \, \lb n - m \rb L_{n+m}
\, + \, \frac{c}{12} \, n \lb n^2 - 1 \rb \delta_{n,-m}
\quad \mathrm{for} \quad \phi_n \! \lb \zeta \, e \rb
\, = \, \frac{L_n}{\zeta^{n+2}} \ , \quad
L_n \, = \, \phi_n \! \lb e \rb
\ . \qquad
\eeq
The point is that the second argument, $z_{12}\,$,
of $T_{2}$ cancels the singular factor
$\frac{1}{z_{12}^{\ 2}}$ in the OPE in the $1$--dimensional
case.

Using the orthogonality of different quasipimary
fields we can produce a sample of projected commutation
relations between $\phi_n \lb z \rb$ for non--collinear
arguments illustrating the appearance of the Virasoro
subalgebra as a special case.

To begin with we note that the vacuum OPE (\ref{r2.3})
remains valid in the $z$--picture provided we set
\beq\label{3.19}
\lb 12 \rb \, = \, \frac{1}{z_{12}^{\ 2}}
\quad (\, \mathrm{implying} \quad \la 12 \ra \, = \,
\frac{c}{2} \, \lb 12 \rb^2 \, = \,
\frac{c}{2} \, \lb z_{12}^{\ 2} \rb^{-2} \ , \quad
z^2 \, = \, \mathbf{z}^2 + z_4^2 \, )
\  \qquad
\eeq
(the singularity at $z_{12}^{\ 2}=0$ being treated as a
limit from the domain
\(\left| z_1^{\ 2} \right| > \left| z_2^{\ 2} \right|\,\)).
Using the knowledge of the generating function for
the Gegenbauer polynomials,
\beq\label{3.20}
\lb \frac{1}{\lb z \! - \! w \rb^2} \rb^{\lambda}
\, = \, \frac{1}{\lb z^2 \rb^{\lambda}} \,
\lb 1 \, - \, 2 \, \widehat{z} \! \cdot \! \widehat{w} \,
\sqrt{\frac{w^2}{z^2}\,}
\, + \, \frac{w^2}{z^2} \rb^{-\lambda}
\, = \, \frac{1}{\lb z^2 \rb^{\lambda}} \,
\Su_{n \, = \, 0}^{\infty} \,
\lb \frac{w^2}{z^2} \rb^{\frac{n}{2}} \,
C_n^{\lambda} \lb \widehat{z} \! \cdot \! \widehat{w} \rb
\ , \qquad
\eeq
and the expressions (\ref{1.7}) and (\ref{r2.2}) for
$2$--, $3$--, and $4$--point correlation functions of
$\phi$ we can write
the term involving the central extension of the Lie
algebra generated by $\phi_n\,$:
\beqa\label{3.21}
&& \!\!\!\!\!\!\!\!
\la 0 \l| \, \phi_2 \,
\Lb \, \phi_n \! \lb z \rb , \phi_{-n} \! \lb w \rb \, \Rb
\phi_{-2} \, \r| 0 \ra \, = \,
\frac{\lb w^2 \rb^{\frac{n}{2} - 1}}{
\lb z^2 \rb^{\frac{n}{2}}} \,
\, \times
\nn && \!\!\!\!\!\!\!\!
\qquad
\quad \times \,
\lvac \, \phi_{2} \,
\LB
C_{n-2}^1 \lb \widehat{z} \! \cdot \! \widehat{w} \rb
\, \phi_0 \! \lb z \rb \, + \,
\frac{w^2}{z^2} \
C_n^1 \lb \widehat{z} \! \cdot \! \widehat{w} \rb
\phi_0 \! \lb w \rb \, + \,
\frac{c}{2 \, z^2} \
C_{n-2}^2 \lb \widehat{z} \! \cdot \! \widehat{w} \rb \RB
\, \phi_{-2} \, \rvac
\nn && \!\!\!\!\!\!\!\!
n \, \geq \, 1 \ , \quad
(\, C_{-1}^{\, \lambda} \, \equiv \, 0 \, )
\ . \qquad
\eeqa
The Virasoro subalgebra (\ref{3.18}) is recovered for
collinear arguments noting the normalization property for
Gegenbauer polynomials:
\beq\label{3.24}
C_n^{\lambda} \lb 1 \rb \, = \,
\lb \!\!\! \begin{array}{c} n + 2 \, \lambda - 1
\\ n \end{array} \!\!\! \rb \, = \,
\frac{\lb 2 \, \lambda \rb_n}{n!} \ , \quad
(\, \frac{2}{\nu +1} \, \Su_{l \, = \, 0}^{\nu} \,
C_{\mu - l}^1 \lb 1 \rb \, = \, 2 \, \mu + 2 - \nu
\, )
\ . \qquad
\eeq
(In particular, Eq. (\ref{3.21})) reproduces (\ref{3.18})
for $n+m=0\,$.)

The Lie algebra $\mathfrak{L}_V$ of the bilocal
field $V$ is much simpler to describe.
The modes $V_{nm}$ of $V$
satisfying (\ref{r4.15})
and the unit operator span
by themselves
an infinite dimensional Lie
algebra:
\beqa\label{3.29}
\!\!\!\!\!\!\!
&&
\Lb \, V_{n_1 n_2} \lb z_1,\, z_2 \rb ,
V_{n_3 n_4} \lb z_3,\, z_4 \rb \, \Rb \, = \,
\nn \!\!\!\!\!\!\! &&
\qquad
= \,
c \, \mathop{\prod}\limits_{j \, = \, 1}^4 \,
\lb z_j^{\, 2} \rb^{- \, \frac{n_j +1}{2}} \,
\LB
C_{ \l| \! n_1 \r| - 1}^1
\lb \widehat{z}_1 \! \cdot \! \widehat{z}_3 \rb \,
C_{ \l| \! n_2 \r| - 1}^1
\lb \widehat{z}_2 \! \cdot \! \widehat{z}_4 \rb \,
\delta_{n_1,-n_3} \, \delta_{n_2,-n_4} \, + \,
\right.
\nn \!\!\!\!\!\!\! &&
\qquad \quad \,
+ \,
\left.
C_{ \l| \! n_1 \r| - 1}^1
\lb \widehat{z}_1 \! \cdot \! \widehat{z}_4 \rb \,
C_{ \l| \! n_2 \r| - 1}^1
\lb \widehat{z}_2 \! \cdot \! \widehat{z}_3 \rb \,
\delta_{n_1,-n_4} \, \delta_{n_2,-n_3}
\RB
\epsilon \left( n_1 \right) \,
\epsilon \left( n_2 \right)
\, + \,
\nn \!\!\!\!\!\!\! &&
\qquad \quad \,
+ \,
\lb z_1^{\, 2} \rb^{- \, \frac{n_1 + 1}{2}}
\lb z_3^{\, 2} \rb^{- \, \frac{n_3 + 1}{2}}
C_{ \l| \! n_1 \r| - 1}^1
\lb \widehat{z}_1 \! \cdot \! \widehat{z}_3 \rb
\epsilon \left( n_1 \right) \,
\delta_{n_1,-n_3} \, V_{n_2 n_4} \lb z_2,\, z_4 \rb \, + \,
\nn \!\!\!\!\!\!\! &&
\qquad \quad \,
+ \,
\lb z_2^{\, 2} \rb^{- \, \frac{n_2 + 1}{2}}
\lb z_3^{\, 2} \rb^{- \, \frac{n_3 + 1}{2}}
C_{ \l| \! n_2 \r| - 1}^1
\lb \widehat{z}_2 \! \cdot \! \widehat{z}_3 \rb
\epsilon \left( n_2 \right) \,
\delta_{n_2,-n_3} \, V_{n_1 n_4} \lb z_1,\, z_4 \rb \, + \,
\nn \!\!\!\!\!\!\! &&
\qquad \quad \,
+ \,
\lb z_1^{\, 2} \rb^{- \, \frac{n_1 + 1}{2}}
\lb z_4^{\, 2} \rb^{- \, \frac{n_4 + 1}{2}}
C_{ \l| \! n_1 \r| - 1}^1
\lb \widehat{z}_1 \! \cdot \! \widehat{z}_4 \rb
\epsilon \left( n_1 \right) \,
\delta_{n_1,-n_4} \, V_{n_2 n_3} \lb z_2,\, z_3 \rb \, + \,
\nn \!\!\!\!\!\!\! &&
\qquad \quad \,
+ \,
\lb z_2^{\, 2} \rb^{- \, \frac{n_2 + 1}{2}}
\lb z_4^{\, 2} \rb^{- \, \frac{n_4 + 1}{2}}
C_{ \l| \! n_2 \r| - 1}^1
\lb \widehat{z}_2 \! \cdot \! \widehat{z}_4 \rb
\epsilon \left( n_2 \right) \,
\delta_{n_2,-n_4} \, V_{n_1 n_3} \lb z_1,\, z_3 \rb
\ . \qquad
\eeqa
It is, in fact, a central
extension of the infinite dimensional real symplectic algebra
$\mathit{sp} \lb \infty,\, \R \rb\,$.
According to (\ref{r4.16}) the $\phi$--modes belong to
this algebra.
The vacuum representation of $\mathfrak{L}_V$
is characterized by
the energy positivity condition (\ref{r4.17}).

The associative algebra of $V_{nm} \lb z,\, w \rb$
contains an
ideal $\Ii_0$ generated by
\beq\label{3.32}
\LB \, V_{n 0} \lb z,\, w \rb \, ( \, = \,
V_{0 n} \lb w,\, z \rb \, ) \ ; \quad n \, \in \, \Z \, \RB
\, (\, \in \, \Ii_0 \, )
\ . \qquad
\eeq
which annihilates all states in the vector space ${\cal H}_V$
spanned by  polynomials in $V_{-n,\, -m}$ ($n,\, m \in \N$)
acting on the vacuum.
Although $\Ii_0$ may well be represented non--trivially in
other sectors of the theory it is natural to work with
the factor algebra $\Bb_V$
in the vacuum sector. Indeed, $\Bb_V$ can be identified as
the operator algebra, generated by the bilocal field $V\,$,
acting (non--trivially) in ${\cal H}_V\,$.
The relative simplicity of the operator algebra $\Bb_V$
in ${\cal H}_V$ stems from the fact that the modes
$V_{nm} \lb z,\, w \rb$ ($n \neq 0 \neq m$) are
(homogeneous) harmonic functions in $z$ and $w\,$--
see (\ref{r4.15}).
It follows from our analysis of the mode space of the
free field $\varphi \lb z \rb$ that $V_{nm} \lb z,\, w \rb$
span a space of dimension $n^2 m^2$ except for the
diagonal, $n=m\,$, for which the symmetry
of $V$ implies that the dimension of the
space is
$\lb\!\!\!\begin{array}{c}n^2+1\\ 2\end{array}\!\!\!\rb\,$.

The modes $V_{nm}$ are eigenvectors of the
\textit{Cartan elements}
\beq\label{3.34}
h_l \, = \, \frac{l}{2 \, \pi^2} \, \mathop{\int}
V_{-l,\, l} \lb u,\, u \rb \, \delta \lb \sqrt{u^2\,} - 1 \rb \,
\mathrm{d}^4 u
\ , \quad (\, u^2 \, = \, \mathbf{u}^2 \, + \, u_4^2 \, )
\ , \quad l \, \in \, \N
\ . \qquad
\eeq
(Parametrizing $u \in \Sr^3$ by
$u=\lb\sin\psi\sin\theta\cos\varphi ,\,\sin\psi\sin\theta
,\,\sin\psi\cos\theta ,\,\cos\psi\rb$
we can replace the volume element
$\delta \lb \sqrt{u^2\,} - 1 \rb \mathrm{d}^4u$ by
$\sin^2\psi$ $\sin\theta$ $\mathrm{d}\psi$
$\mathrm{d}\theta$ $\mathrm{d}\varphi\,$,
$0 \leq \psi \leq \pi\,$, $0 \leq \theta \leq \pi\,$,
$0 \leq \varphi \leq 2\pi\,$; the normalization factor
$\frac{1}{2 \pi^2}$ fixes the integral (of $1$) over $\Sr^3$
to $1\,$.)
We have, in particular,
\beq\label{3.35}
\lb h_l \, - \, \delta_{lm} \, - \, \delta_{ln} \rb
\left. \left.
V_{-n,\, -m} \lb \widehat{z},\, \widehat{w} \rb \, \r| 0 \ra
\, = \, 0
\quad (\, \mathrm{for} \quad n,\, m \, \in \, \N \, )
\ . \qquad
\eeq
In deriving this property one uses the relation
\beq\label{3.36}
\frac{l}{2 \, \pi^2} \, \mathop{\int} \,
C_{l - 1}^1 \lb \widehat{w} \! \cdot \! u \rb \,
C_{n - 1}^1 \lb u \! \cdot \! \widehat{z} \rb \,
\delta \lb \sqrt{u^2\,} - 1 \rb \, \mathrm{d}^4 u \, = \,
\delta_{ln} \,
C_{n - 1}^1 \lb \widehat{w} \! \cdot \! \widehat{z} \rb
\ . \qquad
\eeq
It follows that the conformal Hamiltonian $H$ defined in
(\ref{3.12}) can be written in the form
\beq\label{3.37}
H \, = \, \Su_{l \, = \, 1}^{\infty} \, l \, h_l
\ . \qquad
\eeq




\sectionnew{
Unitary vacuum representations of $\mathfrak{L}_V$}

We begin by introducing an antiinvolution in
$\Bb_V$ and the associated inner product in ${\cal H}\,$.

We define a \textit{star operator} in the algebra of modes
setting
\beq\label{4.1}
V_{n m} \lb z,\, w \rb^{\, *} \, = \,
V_{-m,\, -n} \lb w,\, z \rb \ (\, = \,
V_{-n,\, -m} \lb z,\, w \rb \, )
\quad \mathrm{for} \quad z ,\, w \, \in \, \overline{M}
\ , \qquad
\eeq
so that
$V \lb z,\, w \rb^{\, *} = V \lb z,\, w \rb$.
\vspace{0.2in}

{\it Remark 5.1}
The antiinvolution (\ref{4.1}) involves a correspondence
between homogeneous harmonic functions of degree $n - 1$
and $-n-1\,$. If we write, for $n,\, m \, > \, 0\,$,
\[
V_{-n,\, -m} \lb z,\, w \rb \, = \,
V_{-n,\, -m}^{\, b_1 \, ... \, b_{n-1}, \,
a_1 \, ... \, a_{m-1}} \,
z_{b_1} \, \dots \, z_{b_{n-1}} \,
w_{a_1} \, \dots \, w_{a_{m-1}} \qquad
\]
then we shall have
\[
V_{-n,\, -m} \lb z,\, w \rb^{\, *} \, = \,
V_{m n} \lb w,\, z \rb \, = \,
\frac{1}{w^2 \, z^2} \
V_{m n}^{\, a_1 \, ... \, a_{m-1}, \
b_1 \, ... \, b_{n-1}} \
\frac{w_{a_1}}{w^2} \ \dots \ \frac{w_{a_{m-1}}}{w^2} \
\frac{z_{b_1}}{z^2} \ \dots \ \frac{z_{b_{n-1}}}{z^2}
\ , \qquad
\]
where both $V_{-n,\, -m}$ and $V_{n m}$ are symmetric
traceless tensors of rank $\lb n-1,\, m-1 \rb$
(with respect to the indices $a_i$ and $b_j\,$, separately).
\vspace{0.2in}

We shall call a Hilbert space  (${\cal H}$) representation
of $\mathfrak{L}_V$ \textit{unitary} if the (positive) scalar
product in ${\cal H}$ and the conjugation (\ref{4.1}) in
$\mathfrak{L}_V$ are related by
\beq\label{4.8}
\lb \, \Phi ,\, X \, \Psi \, \rb \, = \,
\lb \, X^* \, \Phi ,\, \Psi \, \rb
\quad \mathrm{for \ every} \quad
\, X \, \in \, \mathfrak{L}_V
\ , \quad \Phi ,\, \Psi \, \in \, {\cal H}^F
\ , \qquad
\eeq
where ${\cal H}^F$ is the dense subspace of finite energy
vectors of ${\cal H}$ which belongs to the domain of any $X$ in
$\mathfrak{L}_V\,$.

One can introduce a (not necessarily positive) inner product
$\la \ , \, \ra$ in ${\cal H}_V$ satisfying (\ref{4.8})
defining the bra vacuum by conditions conjugate to
(\ref{r4.17}):
\beq\label{4.9}
\la 0 \l| V_{n m} \right. \right. \, = \, 0
\quad \mathrm{unless} \quad n \, > \, 0
\quad \mathrm{and} \quad m \, > \, 0
\ , \qquad
\eeq
and assuming $\left. \la 0 \r| 0 \ra = 1\,$.
The main result of this section is the following
characterization of the unitary vacuum representation of
$\mathfrak{L}_V\,$.
\vspace{0.2in}

{\bf Theorem 5.1}
\textit{The inner product in} ${\cal H}_V\,$,
\textit{defined
for a} (\textit{normalized}) \textit{vacuum vector satisfying}
(\ref{r4.17}) \textit{and} (\ref{4.9}) \textit{and for}
$V_{n m} \lb z,\, w \rb$ \textit{obeying} (\ref{3.29}),
\textit{is positive semidefinite iff} $c \in \Z_+$
$= \LB 0,\, 1,\, 2,\, ... \, \RB\,$.
\vspace{0.2in}

{\it Proof.}
Fix a unit vector $e \in \Sr^3$ and consider the
$1$--dimensional
subalgebra $\mathfrak{L}_V^e$ of $\mathfrak{L}_V$ generated by
\beq\label{4.10}
v_{n m} \, := \, V_{n m} \lb e,\, e \rb \ \in \
\mathfrak{L}_V^e \subset \mathfrak{L}_V \ , \quad
n,\, m \, \in \, \Z \ , \quad e^2 \, = \, 1
\ . \qquad
\eeq
It follows from (\ref{3.29}) and from (\ref{3.24}) that
$v_{nm}$ satisfy the commutation relations of the modes
of a $1$--dimensional (chiral) bilocal current:
\beqa\label{4.11}
\Lb \, v_{n_1 m_1} , v_{n_2 m_2} \, \Rb \, =
&& \!\!\!\!\!\!
c \, n_1 \, m_1
\lb \delta_{n_1,\, - n_2} \, \delta_{m_1,\, - m_2}
+ \delta_{n_1,\, - m_2} \, \delta_{m_1,\, - n_2} \rb
\, + \,
\nn && \!\!\!\!\!\! + \,
n_1
\lb \delta_{n_1,\, - n_2} \, v_{m_1 m_2}
+ \delta_{n_1,\, - m_2} \, v_{m_1 n_2} \rb
\, + \,
\nn && \!\!\!\!\!\! + \,
m_1
\lb \delta_{m_1,\, - n_2} \, v_{n_1 m_2}
+ \delta_{m_1,\, - m_2} \, v_{n_1 n_2} \rb
\ . \qquad
\eeqa
\vspace{0.2in}

{\bf Lemma 5.2}
\textit{There is a vector}
$\left. \left. \r| \Delta_n \ra \in
{\cal H}_V^{\lb n \lb n + 1 \rb \rb}$
\textit{whose norm square is a multiple of}
$c \lb c-1 \rb \, \dots \, \lb c-n+1 \rb\,$:
\beqa\label{4.12}
\la \Delta_n \l| \right. \right. \, =
&& \!\!\!\!\!\!\!
\frac{1}{n!} \ \la 0 \l| \right. \right. \!
\left| \!\!
\begin{array}{llll}
v_{11} & v_{12} & \dots & v_{1n} \\
v_{21} & v_{22} & \dots & v_{2n} \\
\dots & \dots & \dots & \dots \\
v_{n1} & v_{n2} & \dots & v_{nn}
\end{array} \!\! \right|
\ , \quad
\nn
\left. \la \Delta_n \r| \Delta_n \ra \, \equiv
\,
\left\| \left. \left. \r| \Delta_n \ra \right\|^2
\, =
&& \!\!\!\!\!\!\!
\lb n+1 \rb ! \
c \lb c-1 \rb \, \dots \, \lb c-n+1 \rb
\ . \qquad {\gvspe}^{\gvspe}
\eeqa
\vspace{0.2in}

{\it Proof.}
It follows from (\ref{4.11}) that the norm square of a
polynomial of degree $n$ in $v_{kl}$ is a polynomial of
degree (not exceeding) $n$ in $c\,$.
We shall demonstrate that
$\left. \la \Delta_n \r| \Delta_n \ra$
vanishes for integer $c$ in the interval $0 \leq c < n\,$.
To this end we note that if $c$ is a positive integer
and $\vec{J}_m\,$, $m \in \Z$ are $c$--dimensional operator
valued vectors
$\vec{J}_m = \LB J_m^i \, , \ i= 1,\, ...,\, c \RB$
satisfying
\beq\label{4.13}
\Lb \, J_m^i \, , \, J_n^j \, \Rb \, = \,
m \, \delta_{m,\, -n} \, \delta_{ij} \ , \quad
m,\, n \, \in \, \Z \ , \quad i,\, j \, = \, 1,\, ...,\, c
\ , \qquad
\eeq
then the normal products
\beq\label{4.14}
v_{lm}^{\lb c \rb} \, = \,
: \! \vec{J}_l \cdot \! \vec{J}_m \! : \ \equiv \
\Su_{i \, = \, 1}^c \, : \! J_l^i \, J_m^i \! :
\qquad
\eeq
satisfy the commutation relations (\ref{4.11}).
If $c < n$ then
$\left. \det \, \lb v_{ij} \rb
\right|_{\, i,\, j \, = \, 1,\, ...,\, n}$
appearing in the definition of
$\la \Delta_n \l| \right. \right.\,$,
which is the Gram determinant of the scalar products of
$n$ vectors in a $c$--dimensional space, should vanish.
The coefficient $\lb n+1 \rb!$ to the leading ($n$th) power
of $c$ is computed as a sum of norm squares of terms
entering the expansion of the determinant;
for instance, for $n = 4$ we have
\beqa\label{no3}
\mathop{\lim}\limits_{c \, \to \, \infty}
\, \lb \frac{1}{c^4} \,
\left. \la \Delta_4 \r| \Delta_4 \ra \rb
\, =
&& \!\!\!\!\!\!\!
\frac{1}{4!^2 \, c^4} \ \LB
{\left\| \la 0 \l| V_{11} \, \dots \, V_{44}
\right. \right. \right\|}^2
\, + \,
6 \, \left\| \la 0 \l| V_{12}^2 \, V_{33} \, V_{44}
\right. \right. \right\|^2
\, + \,
\right.
\nn && \!\!\!\!\!\!\!
\qquad \quad \ + \
4 \, \left\| \la 0 \l| V_{12} \, V_{23} \, V_{13}\, V_{44}
\right. \right. \right\|^2
\, + \,
3 \, \left\| \la 0 \l| V_{12}^2 \, V_{34}^2
\right. \right. \right\|^2
\, + \, \dvspe
\nn && \!\!\!\!\!\!\!
\qquad \quad \ + \, \left.
3 \, \left\|
2 \, \la 0 \l| V_{12} \, V_{23} \, V_{34}\, V_{14}
\right. \right. \right\|^2
\vspe \! \RB \, = \, \vspe
\nn
=
&& \!\!\!\!\!\!\!
2^4 \, + \, 6 \! \times \! 8 \, + \, 4 \! \times \! 8
\, + \, 3 \! \times \! 4 \, + \, 3 \! \times \! 4 \, = \,
120 \, (\, = \, 5! \, ) \gvspe
\ . \qquad
\nonumber
\eeqa
\vspace{0.2in}

{\it Remark 5.2}
The Lie  algebra ${\mathcal L}_V$ of bilocal modes,
characterized by the commutation relations (\ref{3.29})
has a reductive star subalgebra ${\mathcal U}_{\infty}$
(with no central extension) generated by
$V_{-n,\, m} \lb z,\, w \rb\,$, $n,\, m \, \in \, \N\,$:
\beqa\label{4.2}
&& \!\!\!\!\!\!\!
\Lb \, V_{-n_1,\, m_1}
\lb \widehat{z}_1,\, \widehat{w}_1 \rb \, , \,
V_{-n_2,\, m_2}
\lb \widehat{z}_2,\, \widehat{w}_2 \rb \, \Rb \, = \,
\nn && \!\!\!\!\!\!\!
\qquad \qquad
= \, C_{m_1 -1}^1
\lb \widehat{w}_1 \! \cdot \! \widehat{z}_2 \rb \,
\delta_{m_1 n_2} \, V_{-n_1,\, m_2}
\lb \widehat{z}_1,\, \widehat{w}_2 \rb  \, - \,
C_{m_2 -1}^1
\lb \widehat{w}_2 \! \cdot \! \widehat{z}_1 \rb \,
\delta_{m_2 n_1} \, V_{-n_2,\, m_1}
\lb \widehat{z}_2,\, \widehat{w}_1 \rb
\ ,
\nn && \!\!\!\!\!\!\!
(\, \widehat{z}_i^{\, 2} \, = \,
\widehat{w}_j^{\, 2} \, = \, 1 \, )
\eeqa
with a central element
\beq\label{4.3}
{\cal C}_1 \, = \, \Su_{n \, = \, 1}^{\infty} \, h_n
\ , \qquad
\eeq
where $h_n$ are the Cartan operators (\ref{3.34}).
We have
\beqa\label{4.4}
&& \!\!\!\!\!\!\!
\Lb \, V_{-l,\, m}
\lb \widehat{z},\, \widehat{w} \rb ,
{\cal C}_1 \, \Rb \, = \,
\nn && \!\!\!\!\!\!\!
\qquad \qquad
= \, \frac{1}{2 \, \pi^2} \
\mathop{\int} \, \LB \vspe \!
m \, C_{m -1}^1
\lb \widehat{w} \! \cdot \! u \rb \,
V_{-l,\, m}
\lb \widehat{z},\, u \rb  \, - \, \right.
\nn && \!\!\!\!\!\!\!
\qquad \qquad \qquad \qquad \quad \ \,
\left. - \,
l \, C_{l -1}^1
\lb \widehat{z} \! \cdot \! u \rb \,
V_{-l,\, m}
\lb u,\, \widehat{w} \rb \vspe \! \RB
\, \delta \lb \sqrt{u^2\,} - 1 \rb \, \mathrm{d}^4 u \, = \, 0
\ , \quad
\eeqa
where we again used the relation (\ref{3.36}).
${\mathcal U}_{\infty}$ contains what could be
called the \textit{Cartan subalgebra} of
${\mathcal L}_V$ spanned by the elements
$V_{-n,\, n} \lb e,\, e \rb$ for  $n \in \N\,$,
$e^2 = 1$ (including $h_l$ (\ref{3.34})).
${\mathcal L}_V$ is compounded by
${\mathcal U}_{\infty}\,$,
the unit element and by a pair of conjugate abelian
subalgerbas $\mathfrak{L}^{\pm}$
(which are ${\mathcal U}_{\infty}$ modules), spanned by
\beq\label{4.5}
\mathfrak{L}^+ \, \supset \,
\LB \, V_{-n,\, -m} \lb z,\, w \rb \, \RB
\ , \quad
\mathfrak{L}^- \, \supset \,
\LB \, V_{n m} \lb z,\, w \rb \, \RB
\ , \quad n,\, m \, \in \, \N
\ . \qquad
\eeq
$\mathfrak{L}^+$ consists of positive, $\mathfrak{L}^-\,$, of
negative root vectors with respect to the Cartan elements
$h_l$ (\ref{3.34}):
\beq\label{4.6}
\Lb \, h_l \, , \,
V_{\mp n,\, \mp m} \lb \widehat{z}, \, \widehat{w} \rb \, \Rb
\, = \, \pm \lb \delta_{ln} + \delta_{lm} \rb \,
V_{\mp n,\, \mp m} \lb \widehat{z}, \, \widehat{w} \rb
\ . \qquad
\eeq
The commutators between elements of $\mathfrak{L}^-$ and
$\mathfrak{L}^+$
belong to ${\mathcal U}_{\infty} \! \cup c {\mathbf 1}\,$.
The operator
\beq\label{4.7}
{\cal C}_2 \, = \,
\Su_{n,\, m \, = \, 1}^{\infty} \,
\frac{n \, m}{8 \, \pi^4} \
\mathop{\int} \!\!\! \mathop{\int} \,
V_{-n,\, -m} \lb v,\, u \rb \,
V_{m n} \lb u,\, v \rb \,
\, \delta \lb \sqrt{u^2\,} - 1 \rb
\delta \lb \sqrt{v^2\,} - 1 \rb \,
\mathrm{d}^4 u \, \mathrm{d}^4 v
\eeq
commutes with ${\mathcal U}_{\infty}$ and should have a
positive spectrum in any unitary representation of
$\mathfrak{L}_V\,$.
The counterpart of ${\cal C}_2$ (\ref{4.7}) for the
subalgebra $\mathfrak{L}_V^e\,$,
\beq\label{4.15}
{\cal C}_2^e \, = \, \frac{1}{2} \ \Su_{n,\, m \, \geq \, 1}
\, \frac{1}{n \, m} \ v_{-n,\, -m} \, v_{m n}
\qquad
\eeq
has its minimal eigenvalue in the subspace
\[
{\cal H}_e^{\lb n \rb} = \LB \,
\left. \left. P_n \lb v_{-k,\, -l} \rb \r| 0 \ra ; \
P_n \ \mathrm{homogeneous \ of \ degree \ } n
\mathrm{\ in \ } v_{-k, \, -l} \, \RB \qquad
\]
on the vector $\left. \left. \r| \Delta_n \ra$
(conjugate to) (\ref{4.12}):
\beq\label{4.16}
\left. \left. {\cal C}_2^e \, \r| \Delta_n \ra \, = \,
\left. \left. n \lb c - n + 1 \rb \, \r| \Delta_n \ra
\ , \quad
\left.
\Lb \, {\cal C}_2^e - n \lb c - n + 1 \rb \, \Rb \dvspe \!
\right|_{{\cal H}_e^{\lb n \rb}} \, \geq \, 0
\ . \qquad
\eeq
We have, for instance,
\[
\lb {\cal C}_2^e - n \Lb c +2 \lb n - 1 \rb \Rb \rb \,
\left. \left. v_{-k,\, -k}^n \r| 0 \ra \, = \,
0 \, = \,
\lb {\cal C}_2^e - n \, c \rb \,
\left. \left.
v_{-2n,\, -\lb 2n - 1 \rb} \, \dots \, v_{-2,\, -1}
\r| 0 \ra \ . \qquad
\]
\vspace{0.2in}

It follows from Lemma 5.2 that there exist negative norm
vectors unless $c$ is a positive integer.
To prove that for $c \in \N$ the vacuum representation
of $\mathfrak{L}_V$ is indeed unitary it suffices to note that
in this case $V$ can be written in the form
\beq\label{4.17}
V \lb z_1,\, z_2 \rb \, = \, \Su_{i \, = \, 1}^c
\, : \! \varphi_i \lb z_1 \rb \varphi_i \lb z_2 \rb \! :
\ . \qquad
\eeq
where $\varphi_i$ are mutually commuting free zero mass
fields and to recall that a system of free fields satisfies
all Wightman axioms (including positivity).$\quad \Box$
\vspace{0.2in}

We have established on the way the following result
(as a direct consequence of Lemma 5.2).
\vspace{0.2in}

{\bf Proposition 5.3}
\textit{The vacuum representation
of the infinite dimensional Lie algebra}
$\mathfrak{L}_v$
\textit{of the}
$2$--\textit{dimensional} (\textit{2D})
\textit{bilocal chiral field}
\beq\label{4.18}
v \lb z,\, w \rb \, = \, \frac{1}{z\, w} \
\Su_{n,\, m \, \in \, \Z} \, v_{nm} \, z^{-n} \, w^{-m}
\ , \quad z,\, w \, \in \, \C
\qquad
\eeq
\textit{whose modes satisfy} (\ref{4.11})
(\textit{and} $\left. \left. v_{m n} \, \r| 0 \ra = 0$
\textit{unless} $m < 0$ \textit{and} $n < 0\,$)
\textit{is only unitary for positive integer} $c\,$.
\vspace{0.2in}

This is an analogue of Kac--Radul theorem
\cite{KR 93-96} on the unitary representations of the
$W_{1\, +\infty}$ algebra.
It is clear that the algebra of the 2D stress tensor
\beq\label{4.19}
T \lb z \rb \, = \, \frac{1}{2} \ v \lb z,\, z \rb \, = \,
\Su_{n \, \in \, \Z} \, L_n \, z^{-n-2}
\ , \qquad
\eeq
--{\ }i.e. the Virasoro algebra (\ref{3.18}){\ }-- is a true
subalgebra of $\mathfrak{L}_v$ since it admits unitary
representations for all $c \geq 1$ as well as a discrete
series for
$c = c_n = 1 - \, \frac{6}{\lb n+1 \rb \lb n+2 \rb}$
($n = 1,\, 2,\, ... \,$,
the unitary Virasoro module ${\cal H}_{c_n}$
being the quotient space of the corresponding
lowest weight module with respect to
a singular vector at "level" (= eigenvalue of $L_0\,$)
$n \lb n+1 \rb\,$).

The situation is different for $D = 4$ since
$V \lb z,\, w \rb$
is harmonic in each argument in that case.
Due to Corollary 2.4 the algebra $\Bb_V$ is then not bigger
than the original OPE algebra of $\phi$ so that the result
of Theorem 5.1 extends to it.
\vspace{0.2in}

{\bf Corollary 5.4}
\textit{Under the assumptions of Proposition} 2.3
\textit{it follows from Theorem} 5.1
\textit{that the quantum theory of the field}
$\phi$ \textit{with truncated}
$n$--\textit{point function} (\ref{r2.12})
\textit{satisfies Wightman positivity iff} $c$
\textit{is a natural number} (\textit{in which case}
$\phi$ \textit{belongs to the Borchers' class of a
set of free fields}).




\sectionnew{
Extensions of the results.
Concluding remarks}

The preceding results{\ }-- and methods{\ }--
apply to fields of higher dimension and
arbitrary tensor structure.
We shall establish important special cases
of the following
\vspace{0.2in}

\textbf{Conjecture.}
\textit{If a neutral tensor field of integer dimension has}
\textit{truncated}
$n$--\textit{point functions which
are multiples of the corresponding correlators of normal
products of (derivatives of) free fields
for}
$n\leq 6\,$,
\textit{then Wightman
positivity implies that the proportionality constant is
a positive integer.}
(\textit{As indicated in Sec. 2,
for} $d=2$ \textit{the statement follows from the expression
for the $4$--point function}.)

\vspace{0.2in}

Our first example is a conserved current whose
(first five) truncated correlation functions
are obtained from those of the current of a
system of $2$--component spinors,
\beq\label{r6.1}
J^{\mu} \left( x;\, c_{\psi} \right) \, = \,
\Su_{j \, = \, 1}^{c_{\psi}} \,
: \! \psi_j^* \left( x \right)
\, \widetilde{\sigma}^{\, \mu} \,
\psi_j \left( x \right) \! :
\, , \quad
(\, - \widetilde{\sigma}^{\, 0} \, ) \, = \,
\widetilde{\sigma}_0 \, = \, 1\!\!\mathrm{I} \, = \,
\sigma_0
\, , \quad
\widetilde{\sigma}^{\, j} \, = \, - \sigma^j \, = \,
- \sigma_j
\, , \qquad
\eeq
by substituting the positive integer $c_{\psi}$
by an arbitrary real number.
Here $\psi_j$ are mutually anticommuting free Weyl
fields:
\beq\label{r6.2}
\lvac \,
\psi_j \left( x_1 \right) \,
\psi_k^* \left( x_2 \right)
\, \rvac
\, = \,
\delta_{jk} \,
S \left( x_{12} \right)
\, , \quad
S\left( x_{12} \right) \, = \,
i \,
\di \dti_{\ 2} \left( 12 \right)
\, = \,
i \,
\frac{x \dti_{\ 12}}{2 \, \pi^2 \, \rho_{12}^2}
\ , \qquad
\eeq
and we have used the conventions
\beq\label{r6.3}
\di \dti_{\ 2} \, = \,
\sigma_{\mu} \, \frac{\di}{\di x_{2\, \mu}}
\quad
(\, x \dti \, = \, \sigma_{\mu} x^{\mu} \, )
\, , \quad
\sigma^{\mu} \, \wti{\sigma}_{\nu}
\, + \,
\sigma_{\nu} \, \wti{\sigma}^{\mu}
\, = \,
- \, 2 \, \delta^{\mu}_{\, \nu}
\, . \qquad
\eeq
Introducing the spin--tensor components of the current
\beq\label{r6.4}
J \left( x \right)
(\, = \, J_{\alpha\, \dot{\beta}} \left( x \right) \, )
\, = \,
\frac{1}{2} \
\sigma_{\mu} J^{\mu} \,
(\,
= \, \Su_{j \, = \, 1}^{c_{\psi}}
\, :\! \psi_{j\alpha} \left( x \right) \,
\psi^*_{j\dot{\beta}} \left( x \right) \! :
\, )
\, . \qquad
\eeq
we can write
\beqa\label{r6.5}
&& \!\!\!\!\!\!\!\!
\lvac \,
J_{\alpha_1\, \dot{\beta_1}} \left( x_1 \right) \,
J_{\alpha_2\, \dot{\beta_2}} \left( x_2 \right) \,
\rvac
\, = \,
c_{\psi} \,
S_{\alpha_1\, \dot{\beta_2}} \left( x_{12} \right) \
{{}^{{}^{{}^{}}}}^t\!
S_{\dot{\beta_1}\, \alpha_2} \left( x_{12} \right)
\, = \,
\nn && \!\!\!\!\!\!\!\!
\qquad\qquad\qquad\qquad\qquad\qquad
\, = \,
c_{\psi} \,
\LB
S_{\alpha_1\, \dot{\beta_2}} \left( x_{12} \right)
\,
S_{\alpha_2\,\dot{\beta_1}} \left( x_{12} \right)
\, - \,
\frac{\epsilon_{\alpha_1\, \alpha_2}
\epsilon_{\dot{\beta_1}\, \dot{\beta_2}}}{
4 \, \pi^4 \, \rho_{12}^3}
\RB
\, , \qquad
\\ \label{r6.6} && \!\!\!\!\!\!\!\!
J_{\alpha_1\, \dot{\beta_1}} \left( x_1 \right) \,
J_{\alpha_2\, \dot{\beta_2}} \left( x_2 \right)
\, - \,
\lvac \,
J_{\alpha_1\, \dot{\beta_1}} \left( x_1 \right) \,
J_{\alpha_2\, \dot{\beta_2}} \left( x_2 \right) \,
\rvac
\, = \,
{{}^{{}^{{}^{}}}}^t\!
S_{\dot{\beta_1}\, \alpha_2} \left( x_{12} \right) \,
V_{\alpha_1\, \dot{\beta_2}} \left( x_1,\, x_2 \right)
\, + \, \vspe
\nn && \!\!\!\!\!\!\!\!
\qquad\qquad\qquad\qquad\qquad\qquad
\, + \,
S_{\alpha_1\, \dot{\beta_2}} \left( x_{12} \right)\
{{}^{{}^{{}^{}}}}^t\!
V_{\dot{\beta_1}\, \alpha_2} \left( x_1,\, x_2 \right) \,
\, + \,
:\!
J_{\alpha_1\, \dot{\beta_1}} \left( x_1 \right) \,
J_{\alpha_2\, \dot{\beta_2}} \left( x_2 \right) \,
\! :
\, , \qquad
\eeqa
where ${{}^{{}^{{}^{}}}}^t\! S$
($\,{{}^{{}^{{}^{}}}}^t\! V\,$)
stands for the transposed of $S$ ($\, V\,$).
Multiplying both sides by
\(\frac{2\, \pi^2}{i}\, \rho_{12} \,
\wti{x}_{12}^{\, \dot{\beta}_2 \alpha_2}\)
and setting
\beq\label{r6.7}
W_{\alpha_1\, \dot{\beta_1}} \left( x_1,\, x_2 \right)
\, = \,
\frac{2\, \pi^2}{i}\
\rho_{12}\ \wti{x}_{12}^{\, \dot{\beta}_2 \alpha_2}
\, \LB\vspe\!
J_{\alpha_1\, \dot{\beta_1}} \left( x_1 \right) \,
J_{\alpha_2\, \dot{\beta_2}} \left( x_2 \right)
\, - \,
\lvac \,
J_{\alpha_1\, \dot{\beta_1}} \left( x_1 \right) \,
J_{\alpha_2\, \dot{\beta_2}} \left( x_2 \right) \,
\rvac
\vspe\!\RB
\, \qquad
\eeq
we obtain
\beq\label{r6.8}
W_{\alpha_1\, \dot{\beta_1}} \left( x_1,\, x_2 \right)
\, = \,
V_{\alpha_1\, \dot{\beta_1}} \left( x_1,\, x_2 \right)
\, + \,
{{}^{{}^{{}^{}}}}^t\!
V_{\dot{\beta_1}\, \alpha_1} \left( x_1,\, x_2 \right) \,
\, + \,
:\!
J_{\alpha_1\, \dot{\beta_1}} \left( x_1 \right) \,
\mathrm{tr} \lb \wti{x}_{12} \,
J \left( x_2 \right) \rb \,
\! :
\, \qquad
\eeq
where the bilocal field $V$ satisfies
\beq\label{r6.9}
\lvac \,
V_{\alpha_1\, \dot{\beta_1}} \left( x_1,\, x_2 \right) \,
V_{\alpha_2\, \dot{\beta_2}} \left( x_3,\, x_4 \right)
\, \rvac
\, = \,
S_{\alpha_1\, \dot{\beta_2}} \left( x_{14} \right) \
{{}^{{}^{{}^{}}}}^t\!
S_{\dot{\beta_1}\, \alpha_2} \left( x_{23} \right) \,
\, . \qquad
\eeq
It follows from (\ref{r6.9}) that
\beq\label{r6.10}
\wti{\di}^{\ \dot{\alpha} \, \alpha_1}_1 \,
V_{\alpha_1\, \dot{\beta_1}} \left( x_1,\, x_2 \right)
\, = \, 0 \, = \,
V_{\alpha_1\, \dot{\beta_1}} \left( x_1,\, x_2 \right) \,
\frac{\mathop{\di}\limits^{\leftarrow}}{\di x_{2 \, \mu}} \
\wti{\sigma}_{\mu}^{\ \dot{\beta}_1 \, \beta}
\, . \qquad
\eeq
As a result $V_{\alpha\dot{\beta}}$ and the normal
product of $J$ appearing in the right hand side of
(\ref{r6.9}) can be
determined separately and we can prove as in Sec. 5 that
\(c_{\psi}\left(c_{\psi}-1\right)\dots\left(c_{\psi}-n+1\right)
\geq 0\) for \(n=1,2,...\,\).

As a second example we consider the Lagrangean density
\beq\label{r6.11}
{\cal L}_F \left( x \right)
\, = \,
-\, \frac{1}{4} \
\Su_{a \, = \, 1}^{c_F} \
:\!
F_{\mu\nu}^{\ a} \left( x \right) \,
F_{a}^{\mu\nu} \left( x \right)
\! :
\quad (\, c_F \, \in \, \N \, )
\, \qquad
\eeq
and the associated analytic continuation
of truncated Wightman functions to arbitrary
positive real $c_F\,$.
The truncated $n$--point function of ${\cal L}_{F}$
can again be written as a sum of
$\frac{1}{2}\, \left( n-1 \right) !$
$1$--loop graphs, the propagator associated
with a line joining the vertices $1$ and $2$
being
\beqa\label{r6.12}
{\cal D}_{\lambda_1\mu_1\lambda_2\mu_2}
\left( x_{12} \right)
&& \!\!\!\!\!\!
= \,
\frac{1}{4} \
\LB \vspe \!
\partial_{\lambda_1}
\lb \partial_{\lambda_2} \, \eta_{\mu_1 \mu_2}
-
\partial_{\mu_2} \, \eta_{\mu_1 \lambda_2} \rb
\, - \,
\partial_{\mu_1}
\lb \partial_{\lambda_2} \, \eta_{\lambda_1 \mu_2}
-
\partial_{\mu_2} \, \eta_{\lambda_1 \lambda_2} \rb
\vspe \! \RB \,
\frac{1}{4 \, \pi^2 \rho_{12}} \, = \,
\nn && \!\!\!\!\!\!
= \,
\frac{r_{\lambda_1 \lambda_2} \lb x_{12} \rb \,
r_{\mu_1 \mu_2} \lb x_{12} \rb
\, - \,
r_{\lambda_1 \mu_2} \lb x_{12} \rb \,
r_{\mu_1 \lambda_2} \lb x_{12} \rb}{
4 \, \pi^2 \, \rho_{12}^2}
\ . \qquad
\eeqa

This expression for the propagator also enters the OPE
of two $\mathcal{L}$'s
(together with a tensor valued bilocal field):
\beqa\label{5.9}
&& \!\!\!\!\!
\la 0 \l| \right. \right. \!
{\cal L}_F \lb x_1 \rb {\cal L}_F \lb x_2 \rb \, = \,
\la 0 \l| \right. \right. \!
\LB \vspe \!
2 \, c_F \,
{\cal D}^{\, \lambda_1 \, \mu_1 \, \lambda_2 \, \mu_2}
\left( x_{12} \right)
\,
{\cal D}_{\, \lambda_1 \, \mu_1 \, \lambda_2 \, \mu_2}
\left( x_{12} \right)
\, + \,
\right.
\nn && \!\!\!\!\! \qquad\qquad\qquad\qquad\qquad \ \
\left.
+ \,
{\cal D}^{\lambda_1 \, \mu_1 \, \lambda_2 \, \mu_2}
\left( x_{12} \right)
\,
V_{\lambda_1 \, \mu_1 \, \lambda_2 \, \mu_2}
\lb x_1,\, x_2 \rb
\, + \,
: \! {\cal L}_F \lb x_1 \rb {\cal L}_F \lb x_2 \rb \! :
\vspe \! \RB
\ , \quad
\nn && \!\!\!\!\!
2 \,
{\cal D}^{\, \lambda_1 \, \mu_1 \, \lambda_2 \, \mu_2}
\,
{\cal D}_{\, \lambda_1 \, \mu_1 \, \lambda_2 \, \mu_2}
\, = \,
\frac{3}{\lb \pi \, \rho_{12} \rb^4}
\ . \qquad
\eeqa
For $c\in\N\,$, $V$ has a realization as a sum of
normal products of free
Maxwell fields:
\beq\label{5.10}
V_{\lambda_1 \, \mu_1 \, \lambda_2 \, \mu_2}
\lb x_1,\, x_2 \rb \, = \
: \! F_{\lambda_1 \, \mu_1}^{\ a} \lb x_1 \rb
F_{\lambda_2 \, \mu_2}^{\ a} \lb x_2 \rb \! :
\ . \qquad
\eeq

The OPE (\ref{5.9}) allows to compute the truncated
$4$--point function of ${\cal L}_F$ which appears as a
special case of the $5$--parameter expression
$\W_4^{\, t} \left( d=4 \right)$
computed from Eqs. (\ref{1.3})--(\ref{1.6}):
\beqa\label{nn6.15}
{\cal W}_{\, 4}^{\, t} \lb 4 \rb \, = \,
\frac{\rho_{13}^2 \, \rho_{24}^2}{
\rho_{12}^3 \, \rho_{23}^3 \, \rho_{34}^3 \,
\rho_{14}^3} \
&& \!\!\!\!\!\!\!\!\!\!\!
\LB \vspe \!
c_0 \lb 1 + \eta_1^5 + \eta_2^5 \rb
+ c_1
\lb \eta_1 + \eta_2 + \eta_1^4 + \eta_2^4
+ \eta_1 \, \eta_2 \,
\lb \eta_1^3 + \eta_2^3 \rb \rb + \right.
\nn && \!\!\!\!\!\!\!
+ \ c_2 \lb \eta_1^2 + \eta_2^2 +
\eta_1^3 + \eta_2^3 +
\eta_1^2 \, \eta_2^2 \, \lb \eta_1 + \eta_2 \rb \rb
+
\nn && \!\!\!\!\!\!\!
\left.
+ \ b_1 \, \eta_1 \, \eta_2 \lb 1 +
\eta_1^2 + \eta_2^2 \rb +
b_2 \, \eta_1 \, \eta_2 \lb \eta_1 \, \eta_2 +
\eta_1 + \eta_2 \rb
\vspe \! \RB
\nn && \!\!\!\!\!\!\!
\!\!\!\!\!\!\!\!\!\!\!\!\!\!\!\!\!\!\!\!\!\!\!\!\!\!\!
\!\!\!\!\!\!\!\!\!\!\!\!\!\!\!\!\!\!\!\!\!\!\!\!\!\!\!
\!\!\!\!
(\,
c_i \, \equiv \, c_{0i} \quad \mathrm{for} \quad
i \, = \, 0,\, 1,\, 2 \, ; \quad
b_i \, \equiv \, c_{1i} \quad \mathrm{for} \quad
i \, = \, 1,\, 2
\, )
\ . \qquad \gvspe
\eeqa
Indeed the
contribution $\WW_{\Box}$ of the box diagram
(computed by using formulae for traces of products of
$r_{\ \nu}^{\mu}$ given in Appendix B),
\beqa\label{r6.15}
\WW_{\Box} \, = && \!\!\!\!\!\!\!\!
c_F \,
\DD^{\lambda_1\,\lambda_2}_{\ \ \,\mu_1\mu_2}
\left( x_{12} \right) \,
\DD_{\lambda_1\,\lambda_4}^{\ \ \,\mu_1\mu_4}
\left( x_{14} \right) \,
\DD_{\lambda_2\,\lambda_3}^{\ \ \,\mu_2\mu_3}
\left( x_{23} \right) \,
\DD^{\lambda_3\,\lambda_4}_{\ \ \,\mu_3\mu_4}
\left( x_{34} \right)
\, = \, \dvspe
\nn = && \!\!\!\!\!\!\!\!
32 \, c_F \,
\frac{
\left( 12 \right)
\left( 23 \right)
\left( 34 \right)
\left( 14 \right)}{
\rho_{12}
\rho_{23}
\rho_{34}
\rho_{14}}\
\left( 1 +
\frac{\eta_1}{\eta_2} \ +
\frac{\eta_2}{\eta_1} \ +
\frac{\eta_1}{\eta_2} \ -
\frac{2}{\eta_2} \ -
\frac{2}{\eta_1} \ +
\frac{2}{\eta_1\eta_2} \right)
\, , \qquad
\eeqa
which enters the expression for the truncated
$4$--point function of $\LL \left( x \right)$
\beqa\label{r6.16}
\WW_4^{\ t} \, = && \!\!\!\!\!\!\!\!
\left( 1 + s_{12} + s_{23} \right) \,
\WW_{\Box} \left( x_1,\, x_2,\, x_3,\, x_4 \right)
\, = \,
\nn = && \!\!\!\!\!\!\!\!
\WW_{\Box} \left( x_1,\, x_2,\, x_3,\, x_4 \right)
+
\WW_{\Box} \left( x_2,\, x_1,\, x_3,\, x_4 \right)
+
\WW_{\Box} \left( x_1,\, x_3,\, x_2,\, x_4 \right)
\, \qquad
\eeqa
fits the expression (\ref{nn6.15}) for
\beq\label{r6.17}
c_0 \, = \, c_2 \, = \, b_1 \, = \,
- \frac{1}{2} \ c_1 \, = \, \frac{c_F}{8 \, \pi^8}\,
\, , \quad
b_2 \, = \, 0
\, . \qquad
\eeq
The first local field in the expansion of $V$ around
the diagonal is the stress energy tensor:
\beq\label{r6.18}
T^{\mu}_{\ \nu} \, = \,
\frac{1}{4} \
V^{\kappa\lambda}_{\ \ \, \kappa\lambda} \left( x,\, x \right)
\delta^{\mu}_{\ \nu} -
V^{\lambda\mu}_{\ \ \, \lambda\nu} \left( x,\, x \right)
\, = \,
- \, \LL \left( x \right) \delta^{\mu}_{\nu} -
V^{\lambda\mu}_{\ \ \, \lambda\nu} \left( x,\, x \right)
\, . \qquad
\eeq
Conversely, the bilocal tensor field
\(
V^{\lambda_1\mu_1}_{\ \ \, \lambda_2\mu_2}
\left( x_1,\, x_2 \right)
\, ( \, = \, - V^{\mu_1\lambda_1}_{\ \ \, \lambda_2\mu_2}
\left( x_1,\, x_2 \right) \, = \,
- V^{\lambda_1\mu_1}_{\ \ \, \mu_2\lambda_2}
\left( x_1,\, x_2 \right) \, )
\)
appears in the OPE of two $T^{\mu}_{\, \nu}$
and can be determined from it in two steps.
First, one derives the formula
\beqa\label{r6.19}
\lvac\,
V^{\lambda_1\mu_1}_{\ \ \, \lambda_2\mu_2}
\left( x_1,\, x_2 \right)
\,
V^{\lambda_3\mu_3}_{\ \ \, \lambda_4\mu_4}
\left( x_3,\, x_4 \right)
\,\rvac
\, = && \!\!\!\!\!\!\!\!
c_F \,
\DD^{\lambda_1\mu_1\lambda_3\mu_3} \left( x_{13} \right)
\DD_{\lambda_2\mu_2\lambda_4\mu_4} \left( x_{24} \right)
+
\nn && \!\!\!\!\!\!\!\! \, +
c_F \,
\DD^{\lambda_1\mu_1}_{\ \ \,\lambda_4\mu_4}
\left( x_{14} \right)
\DD^{\ \ \,\lambda_3\mu_3}_{\lambda_2\mu_2}
\left( x_{23} \right)
\, . \qquad
\eeqa
and deduces from it that
\(
V^{\lambda_1\mu_1}_{\ \ \,\lambda_2\mu_2}
\left( x_1,\, x_2 \right)
\)
satisfies in each argument the free Maxwell
equations.
Secondly, one uses this fact
to single out the contribution of $V$ in
the OPE of two $T$'s.
Once more Wightman positivity implies
$c_F \in \N\,$.
\vspace{0.2in}

\textit{Remark 6.1}
The use of different notation,
\(c \, (\, = \, c_{\phi})\,\)
$c_{\psi}$ and $c_F$ for the constants
multiplying the truncated functions of normal
products of the free fields $\phi\,$,
$\psi$ and $F_{\mu\nu}\,$,
respectively, is justified by the fact that
they correspond to (and exhaust the) different
tensor structures in the general conformal
invariant $3$--point function of the stress
energy tensor \cite{Sta88}.
\vspace{0.2in}

At the same time the $4$--point functions of the
conserved current $J_{\mu}$  and
$\LL \left( x \right)$ involve structures
which cannot be reduced to normal products of free
fields.
If, for instance, $b_2 \neq 0$ in (\ref{nn6.15})
the $3$--point function of $\LL \left( x \right)$
won't vanish (unlike the case of superposition
of type (\ref{r6.11}) of normal products of free
Maxwell fields).
More generally, we have a $4$--parameter family
of admissible $4$--point functions of
$\LL \left( x \right)$
obtained from Eq. (\ref{nn6.15}) with the restriction
\beq\label{1.11}
c_2 \, = \, - c_0 \, - \, c_1 \, (\, \neq \, 2 \, c_0 \,)
\  \qquad
\eeq
coming from the requirement that no $d = 2$ field
appears in the OPE of
\({\cal L} \left( x_1 \right){\cal L} \left( x_2 \right)\)
(and that the stress energy tensor is present in this OPE).
They are only compatible with $3$--point functions of
$T$ of the type (\ref{r6.18}) (i.e. with the third of
the three admissible structures in this $3$--point
function given in \cite{Sta88}-- cf. Remark 6.1).

To summarize: looking for a $4$--dimensional RCFT
beyond the Borchers' class of free fields we have
excluded the theory of a bilocal field of dimension
$\left( 1,\, 1 \right)$ and have come to the following
problem.
Assume that the only local fields in the observable
algebra, satisfying GCI, of dimension $d \leq 4$
are the (conserved traceless) stress energy tensor
$T_{\mu\nu} \left( x \right)$
and a scalar field $\LL \left( x \right)$
of dimension $4$ (playing the role of an
action density).
The problem is to construct an OPE algebra
consistent with the $n$--point functions of these
fields for $n \leq 4$ that would allow to
compute higher point correlation functions and to
implement the condition of Wightman positivity.
This example is attractive because
the dimensions of the basic fields ${\cal L}$
and $T_{\mu\nu}$ are protected.
Moreover, in any renormalizable quantum field theory
one can define a (gauge invariant) local action
density and a stress energy tensor.




\pagebreak 

\vspace{0.4in}
\noindent
{\bf\Large Acknowledgments} \\

We thank for support and hospitality the
Erwin Schr\"{o}dinger International Institute
for Mathematical Physics (ESI) in Vienna
where a major part of this work was done.
Discussions with Hans Borchers, Harald Grosse,
Jakob Yngvason and Yuri Neretin are gratefully acknowledged.
We thank a referee for his careful reading of the manuscript
and for useful suggestions.
The research of Ya.{\,}S. was supported in part
by the EEC contracts HPRN-CT-2000-00122
and HPRN-CT-2000-00148 and by the INTAS contract 99-1-590.
N.{\,}N. and I.{\,}T. acknowledge partial support by the
Bulgarian National Council for Scientific Research under
contract F-828.





\appendiX[Appendix~]
\link{equation}{section}\toheight1


\sectionnew{
Proof of Proposition 3.1
}
\label{app:A}

We shall first compute the sum in the right hand side
of (\ref{2.20}) for
\beqa\label{A.1}
&&
\rho_{34} \, = \, 0 \ , \quad
2 \, X_y \! \cdot \! x_{34} \, = \,
\frac{\rho_{14} - \rho_{13}}{\lb 1 - \alpha \rb
\rho_{14} + \alpha \, \rho_{13}} \ - \,
\frac{\rho_{24} - \rho_{23}}{\lb 1 - \alpha \rb
\rho_{24} + \alpha \, \rho_{23}} \ = \,
\nn &&
= \, \frac{\rho_{14} \, \rho_{23} -
\rho_{13} \, \rho_{24}}{
\Lb \lb 1 - \alpha \rb \rho_{14} + \alpha \, \rho_{13} \Rb
\Lb \lb 1 - \alpha \rb \rho_{24} + \alpha \, \rho_{23} \Rb}
\ = \, \frac{- \epsilon}{
\Lb  \alpha + \lb 1 - \alpha \rb
\frac{\rho_{14}}{\rho_{13}} \Rb
\Lb 1 - \alpha + \alpha \,
\frac{\rho_{23}}{\rho_{24}} \Rb} \ , \quad
\nn &&
\frac{\rho_{13} \, \rho_{24}}{\rho_{12}} \ X_y^2
\, = \,  \frac{1}{
\Lb  \alpha + \lb 1 - \alpha \rb
\frac{\rho_{14}}{\rho_{13}} \Rb
\Lb 1 - \alpha + \alpha \,
\frac{\rho_{23}}{\rho_{24}} \Rb}
\ , \qquad
\eeqa
and then use the result to give a general proof of
Proposition 3.1.
According to (\ref{A.1}) we have
\beqa\label{no4}
\mathop{\lim}\limits_{\rho_{34} \to 0} \,
\lb \rho_{34} X_y^2 \rb^l \,
C_{2l}^1 \! \lb
\widehat{X}_y \! \cdot \! \widehat{x}_{34} \rb
\, = \,
\frac{\epsilon^{2l}}{
\Lb  \alpha + \lb 1 - \alpha \rb
\frac{\rho_{14}}{\rho_{13}} \Rb^{2l}
\Lb 1 - \alpha + \alpha \,
\frac{\rho_{23}}{\rho_{24}} \Rb^{2l}}
\ . \qquad \nonumber
\eeqa
Conformal invariance allows to send $x_1$ to infinity
setting $\frac{\rho_{14}}{\rho_{13}} \, \to 1\,$,
$\frac{\rho_{23}}{\rho_{24}} \, \to 1 - \epsilon$
thus reproducing the right hand side of (\ref{2.23}).
Taking the sum in $l$ we reduce the proof of (\ref{2.23})
to verifying the identity
\beq\label{A.2}
2 \mathop{\int}\limits_{\!\!\!\!\!\! 0}^{\;\;\ 1}
\frac{
\lb 1 - \epsilon \, \alpha \rb
\Lb \lb 1 - \epsilon \, \alpha \rb^2 +
\epsilon^2 \alpha^2 \lb 1 - \alpha \rb^2 \Rb }{
\lb 1 - \epsilon \, \alpha^2 \rb
\lb 1 - 2 \, \epsilon \, \alpha +
\epsilon \, \alpha^2 \rb}
\ \, \mathrm{d}\alpha \, = \,
1 \, + \, \frac{1}{1 \! - \! \epsilon}
\qquad
\eeq
which is straightforward.

It is also instructive to compute the individual terms
in the right hand side of (\ref{2.23}) which correspond
to the contribution of twist $2$ fields to the OPE.
Using Euler's integral representation for the
hypergeometric function we find
\beq\label{A.3}
1 \, + \, \frac{1}{1 \! - \! \epsilon} \, = \,
2 \, \Su_{l \, = \, 0}^{\infty} \,
\lb \!\!\!
\begin{array} {l} 4 l \\ 2 l \end{array}
\!\! \rb^{\! -1} \!
\epsilon^{2l} \, F \lb 2l+1,\, 2l+1;\, 4l+2;\, \epsilon \rb
\ . \qquad
\eeq
Each $F \lb 2l+1,\, 2l+1;\, 4l+2;\, \epsilon \rb$ is,
in fact, an elementary function.
In particular, the first two terms which provide the
contribution of the original field $\phi$ and of the
stress--energy tensor $T_2$ to the OPE can be written
in the form
\beqa\label{A.4}
&& \!\!\!\!\!\!\!
\left.
2 \mathop{\int}\limits_{\!\!\!\!\!\! 0}^{\;\;\ 1}
\frac{
\la 0 \l| V \lb x_1,\, x_2 \rb
\phi \lb x_4 + \alpha \, x_{34} \rb
\r| 0 \ra}{
c \lb 13 \rb \lb 24 \rb} \ \, \mathrm{d}\alpha \
\right|
\mathop{\mathop{}\limits_{}}\limits_{\rho_{34} \, = \, 0}
\, = \,
2 \, F \lb 1,\, 1;\, 2;\, \epsilon \rb \, = \,
\nn && \!\!\!\!\!\!\! \qquad
= \, \frac{2}{\epsilon} \ \ln \frac{1}{1 \! - \! \epsilon}
\, = \, 2 \, + \, \epsilon \, + \,
\Su_{n \, = \, 2}^{\infty} \,
\frac{2 \, \epsilon^n}{n \! + \! 1}
\ , \qquad
\nn && \!\!\!\!\!\!\!
\left.
2 \, C_1 \mathop{\int}\limits_{\!\!\!\!\!\! 0}^{\;\;\ 1}
\frac{
\la 0 \l| V \lb x_1,\, x_2 \rb
T_2 \lb x_4 + \alpha \, x_{34},\, x_{34} \rb
\r| 0 \ra}{
c \lb 13 \rb \lb 24 \rb} \ \, \mathrm{d}\alpha \
\right|
\mathop{\mathop{}\limits_{}}\limits_{\rho_{34} \, = \, 0}
\, = \,
\frac{\epsilon^2}{3} \ F \lb 3,\, 3;\, 6;\, \epsilon \rb
\, = \,
\nn && \!\!\!\!\!\!\! \qquad
= \, \frac{60}{\epsilon^2} \,
\Lb
\lb \frac{1}{\epsilon} - 1 + \frac{\epsilon}{6} \rb
\ln \frac{1}{1 \! - \! \epsilon} \, - \,
1 \, + \, \frac{\epsilon}{2}
\Rb
\, = \,
\epsilon^2 \, \LB
\frac{1}{3} \, + \, \frac{\epsilon}{2} \, + \,
\Su_{n \, = \, 2}^{\infty} \,
\frac{\lb 4 \rb_{n-1} \lb 5 \rb_{n-2} \, \epsilon^n}{
\lb 3 \rb_{n-2} \lb 7 \rb_{n-1}}
\RB
\ .
\nn && \!\!\!\!\!\!\!
\qquad \quad
\eeqa

Proceeding to the general case ($\rho_{34} \neq 0$)
we shall use the following generalization of (\ref{A.3})
(see \cite{DO 01}). Exchange the conformal cross ratios
(\ref{1.4})
(\ref{2.21}) with the variables $\eta$ end
$\overline{\eta}$ related to $\eta_1$ and $\epsilon$ by
\beq\label{A.5}
\eta \, \overline{\eta} \ = \eta_1 \ , \quad
\eta \, + \, \overline{\eta} = \epsilon \, + \, \eta_1
\ , \quad (\, \lb 1 - \eta \rb \lb 1 - \overline{\eta} \rb
\, = \, \eta_2 \, )
\quad . \qquad
\eeq
(We note that for space--like $x_{ij}$ the variables
$\eta$ and $\overline{\eta}$ are complex conjugate to
each other.)
In terms  of these variables we can write (see Eq. (3.10)
of \cite{DO 01}):
\beqa\label{A.6}
&& \!\!\!\!\!\!\!\!\!\!\!\!
\frac{2}{X^2_y} \, \lb 4l + 1 \rb
\mathop{\int}\limits_{\!\!\!\!\!\! 0}^{\;\;\ 1}
\mathrm{d}\alpha \ \alpha^{2l} \lb 1 - \alpha \rb^{2l} \,
\Su_{n \, = \, 0}^{\infty} \,
\frac{\lb - \, \frac{\alpha \, \lb 1 - \alpha \rb}{4}
\, \rho_{34} \, \Box_4 \rb^n}{n! \, \lb 2l +1 \rb_n }
\
\rho_{34}^{\, l} \ \lb X_y^2 \rb^{l + 1} \
C_{2l}^1 \! \lb
\widehat{X}_y \! \cdot \! \widehat{x}_{34} \rb
\, = \,
\nn && \!\!\!\!\!\!\!\!\!\!\!\!
= \, 2 \, \lb \!\!\!
\begin{array} {l} 4 l \\ 2 l \end{array}
\!\! \rb^{\! -1}
\frac{\eta^{2l+1} \ F \lb 2l+1,\, 2l+1;\, 4l+2; \eta \rb
\, - \,
\overline{\eta}^{2l+1} \ F \lb 2l+1,\, 2l+1;\, 4l+2;
\overline{\eta} \rb}{\eta \! - \! \overline{\eta}}
\ . \qquad \quad
\eeqa
We can sum up these expressions applying (\ref{A.3});
as a result the $\eta_1$--dependent terms present for each
$l$ cancel and we end up with
\beqa\label{A.7}
&& \!\!\!\!\!\!\!\!\!\!
\frac{2}{\eta \! - \! \overline{\eta}} \, \times
\nn && \!\!\!\!\!\!\!\!\!\!
\ \times \,
\Su_{l \, = \, 0}^{\infty} \,
\lb \!\!\!
\begin{array} {l} 4 l \\ 2 l \end{array}
\!\! \rb^{\! -1} \!
\LB
\eta^{2l+1} \ F \lb 2l+1,\, 2l+1;\, 4l+2; \eta \rb
\, - \,
\overline{\eta}^{2l+1} \ F \lb 2l+1,\, 2l+1;\, 4l+2;
\overline{\eta} \rb
\RB \, = \,
\nn && \!\!\!\!\!\!\!\!\!\!
= \, \frac{1}{\eta \! - \! \overline{\eta}} \
\lb \eta \, + \, \frac{\eta}{1 \! - \! \eta} \, - \,
\overline{\eta} \, - \,
\frac{\overline{\eta}}{1 \! - \! \overline{\eta}} \rb
\, = \,
1 \, + \, \frac{1}{\lb 1 \! - \! \eta \rb
\lb 1 \! - \! \overline{\eta} \rb}
\, = \,
1 \, + \,  \frac{1}{1 \! - \! \epsilon}
\ . \qquad \quad
\eeqa
This completes the proof of Proposition 3.1.$\quad \Box$




\sectionnew{
Traces of products of $r^{\mu}_{\ \nu} \lb x \rb$}
\label{app:C}

We shall compute the trace of the product of tensor
structures that appears in the numerator of the box
diagram with propagator (\ref{r6.12}):
\beqa\label{C.1}
&&
B \, = \, f^{\lambda_1 \mu_1}_{\ \ \lambda_2 \mu_2}
\lb x_{12} \rb \,
f^{\lambda_2 \mu_2}_{\ \ \lambda_3 \mu_3}
\lb x_{23} \rb \,
f^{\lambda_3 \mu_3}_{\ \ \lambda_4 \mu_4}
\lb x_{34} \rb \,
f^{\lambda_4 \mu_4}_{\ \ \lambda_1 \mu_1}
\lb x_{14} \rb \ , \quad
\nn &&
f^{\lambda \mu}_{\ \ \lambda' \mu'}
\lb x \rb \, = \,
r^{\lambda}_{\ \lambda'} \lb x \rb \,
r^{\mu}_{\ \mu'} \lb x \rb
\, - \, r^{\lambda}_{\ \mu'} \lb x \rb \,
r^{\mu}_{\ \lambda'} \lb x \rb
\ , \qquad
\eeqa
establishing on the way some useful properties of
products of \(r^{\mu}_{\ \nu} \lb x \rb
= \delta^{\mu}_{\ \nu} - 2 \ \frac{x^{\mu}\, x_{\nu}}{
x^2 \, + \, i \, 0 \, x^0}\)
(\ref{2.16})
(of different arguments) which appear in correlation
functions of tensor fields.

We shall use repeatedly the triple product formula
of \cite{OP94}:
\beqa\label{C.2}
&&
r \lb x_{12} \rb \, r \lb x_{23} \rb \, r \lb x_{13} \rb
\, = \,
r \lb X_{23} \rb
\ , \quad
\mathrm{i.e.} \quad
\nn &&
r^{\lambda}_{\ \sigma} \lb x_{12} \rb \,
r^{\sigma}_{\ \tau} \lb x_{23} \rb \,
r^{\tau}_{\ \mu} \lb x_{13} \rb
\, = \,
r^{\lambda}_{\ \mu} \lb X_{23} \rb
\ , \quad
X_{23} \, = \, \frac{x_{13}}{\rho_{13}} \ - \,
\frac{x_{12}}{\rho_{12}}
\ . \qquad
\eeqa
Using the identity $r \lb x \rb^2 = {\bf 1}$ we find
\beqa\label{C.3}
R \lb x_{12},\, x_{23},\, x_{34},\, x_{14} \rb
&& \!\!\!\!\!\!\!
:= \,
r \lb x_{12} \rb \, r \lb x_{23} \rb \,
r \lb x_{34} \rb \, r \lb x_{14} \rb
\, = \,
\nn && \!\!\!\!\!\!\!
= \,
\Lb \vspe \!
r \lb x_{12} \rb \, r \lb x_{23} \rb \, r \lb x_{13} \rb \,
\vspe \! \Rb
\Lb \vspe \!
r \lb x_{13} \rb \, r \lb x_{34} \rb \, r \lb x_{14} \rb \,
\vspe \! \Rb
\, = \,
\nn && \!\!\!\!\!\!\!
= \,
r \lb X_{23} \rb \, r \lb X_{34} \rb
\ , \qquad
\eeqa
where
$X_{34}\equiv X_{34}^1 = \frac{x_{14}}{\rho_{14}} -
\frac{x_{13}}{\rho_{13}}$
(cf. (\ref{2.16})).
Using further the relation
\beq\label{C.4}
\mathrm{tr} \lb \vspe \!
r \lb x \rb \, r \lb y \rb
\vspe \! \rb \, (\, = \,
4 \, \frac{\lb x \! \cdot \! y \rb^2}{x^2 \, y^2}
\, + \, D \, - \, 4 \ ) \, = \,
4 \, \frac{\lb x \! \cdot \! y \rb^2}{x^2 \, y^2}
\quad \mathrm{for} \quad D \, = \, 4
\qquad
\eeq
we deduce (for $\eta_i$ given by (\ref{1.4}))
\beq\label{C.5}
\mathrm{tr}
\lb \vspe \!
R \lb x_{12},\, x_{23},\, x_{34},\, x_{14} \rb
\vspe \! \rb
\, = \,
\frac{\lb 2 \, X_{23} \! \cdot \! X_{34} \rb^2}{
X_{23}^{\ 2} \, X_{34}^{\ 2}}
\, = \,
\frac{\lb 1 \! - \! \eta_1 \! - \! \eta_2 \rb^2}{
\eta_1 \, \eta_2}
\ . \qquad
\eeq
A simple algebra allows to reduce $B$ (\ref{C.1}) to
the difference
\beq\label{C.6}
B \, = \,
8 \, \LB
\Lb \vspe \! \,
\mathrm{tr} \,
R \lb x_{12},\, x_{23},\, x_{34},\, x_{14} \rb
\vspe \! \Rb^2
\, - \,
\mathrm{tr} \lb \vspe \! \,
\Lb
\, R \lb x_{12},\, x_{23},\, x_{34},\, x_{14} \rb \,
\Rb^2
\vspe \! \rb
\RB
\ . \qquad
\eeq
The second term is computed using once more (\ref{C.3}):
\beqa\label{C.7}
\mathrm{tr} \lb \vspe \! \,
\Lb
\, R \lb x_{12},\, x_{23},\, x_{34},\, x_{14} \rb \,
\Rb^2
\vspe \! \rb
&& \!\!\!\!\!\!\!
= \,
\mathrm{tr} \lb \vspe \! \,
r \lb X_{23} \rb \, r \lb X_{34} \rb \,
r \lb X_{23} \rb \, r \lb X_{34} \rb \,
\vspe \! \rb
\, = \,
\nn && \!\!\!\!\!\!\!
= \,
\mathrm{tr} \LB \vspe \! \,
r \lb
\frac{\rho_{12} \, x_{14} - \rho_{14} \, x_{12}}{
\rho_{24}^2}
\ + \,
\frac{\rho_{12} \, x_{13} - \rho_{13} \, x_{12}}{
\rho_{23}^2}
\rb
\, \times \right.
\nn && \!\!\!\!\!\!\!
\qquad \quad
\left. \times \,
r \lb
\frac{\rho_{13} \, x_{14} - \rho_{14} \, x_{13}}{
\rho_{34}^2}
\ + \,
\frac{\rho_{12} \, x_{14} - \rho_{14} \, x_{12}}{
\rho_{24}^2}
\rb
\vspe \! \RB
\, = \,
\nn && \!\!\!\!\!\!\!
= \,
\frac{\lb
1 - 2 \, \eta_1  - 2 \, \eta_2 + \eta_1^2 + \eta_2^2
\rb^2}{
\eta_1^2 \, \eta_2^2}
\ . \qquad
\eeqa
Inserting finally (\ref{C.5}) and (\ref{C.7}) we find
\beq\label{C.8}
B \, = \,
\frac{32}{\eta_1 \, \eta_2}
\ \lb
1 - 2 \, \eta_1  - 2 \, \eta_2 + \eta_1^2 + \eta_2^2 +
\eta_1 \, \eta_2
\rb
\ . \qquad
\eeq







{\small
\pagebreak 

}


\end{document}